\PassOptionsToPackage{super,sort&compress}{natbib}
\documentclass[pdflatex,sn-nature]{sn-jnl}
\usepackage{natbib}
\usepackage{xcolor}
\definecolor{lightblue}{RGB}{0,114,178}

\hypersetup{
  colorlinks=true,
  linkcolor=lightblue,
  citecolor=lightblue,
  urlcolor=lightblue
}

\usepackage{graphicx}%
\usepackage{multirow}%
\usepackage{amsmath,amssymb,amsfonts}%
\usepackage{amsthm}%
\usepackage{mathrsfs}%
\usepackage[title]{appendix}%
\usepackage{textcomp}%
\usepackage{manyfoot}%
\usepackage{booktabs}%
\usepackage{algorithm}
\usepackage{algorithmic}
\usepackage{listings}%
\usepackage{changes}
\usepackage{caption}
\usepackage{newfloat}
\DeclareFloatingEnvironment[
    name=\textbf{Supplementary Figure},
    placement=htbp
]{suppfigure}

\DeclareFloatingEnvironment[
    name=\textbf{Supplementary Table},
    placement=htbp
]{supptable}

\DeclareCaptionLabelFormat{boldlabel}{\textbf{#1~#2}}
\DeclareCaptionLabelSeparator{bar}{\space\textbf{$\vert$}\space}
\begin{document}

\title[Article Title]{Generative Modeling Enables Molecular Structure Retrieval from Coulomb Explosion Imaging}

\author*[1]{\fnm{Xiang} \sur{Li}}\email{\href{mailto:xiangli@slac.stanford.edu}{\textcolor{black}{xiangli@slac.stanford.edu}}}

\author[2,3]{\fnm{Till} \sur{Jahnke}}

\author[2]{\fnm{Rebecca} \sur{Boll}}

\author[4]{\fnm{Jiaqi} \sur{Han}}
\author[4]{\fnm{Minkai} \sur{Xu}}

\author[2]{\fnm{Michael} \sur{Meyer}}

\author[5]{\fnm{Maria Novella} \sur{Piancastelli}}

\author[6]{\fnm{Daniel} \sur{Rolles}}
\author[6]{\fnm{Artem} \sur{Rudenko}}

\author[7]{\fnm{Florian} \sur{Trinter}}
\author[1,8]{\fnm{Thomas J.A.} \sur{Wolf}}

\author[1]{\fnm{Jana B.} \sur{Thayer}}
\author[1,8]{\fnm{James P.} \sur{Cryan}}
\author[4]{\fnm{Stefano} \sur{Ermon}}
\author[9]{\fnm{Phay J.} \sur{Ho}}

\affil[1]{\orgdiv{Linac Coherent Light Source}, \orgname{SLAC National Accelerator Laboratory}, \orgaddress{\city{Menlo Park}, \state{CA}, \country{USA}}}

\affil[2]{\orgname{European XFEL}, \orgaddress{\street{Holzkoppel 4}, \city{Schenefeld}, \country{Germany}}}

\affil[3]{\orgname{Max-Planck-Institut für Kernphysik}, \orgaddress{\city{Heidelberg}, \country{Germany}}}

\affil[4]{\orgdiv{Department of Computer Science}, \orgname{Stanford University}, \orgaddress{\city{Stanford}, \state{CA}, \country{USA}}}

\affil[5]{\orgname{Sorbonne Universit\'e, CNRS}, \orgdiv{ Laboratoire de Chimie Physique-Mati\`ere et Rayonnement, LCPMR}, \orgaddress{\city{Paris}, \country{France}}}

\affil[6]{\orgdiv{J. R. Macdonald Laboratory, Department of Physics}, \orgname{Kansas State University}, \orgaddress{\city{Manhattan}, \state{Kansas}, \country{USA}}}

\affil[7]{\orgdiv{Molecular Physics}, \orgname{Fritz-Haber-Institut der Max-Planck-Gesellschaft}, \orgaddress{\city{Berlin}, \country{Germany}}}

\affil[8]{\orgdiv{Stanford PULSE Institute}, \orgname{SLAC National Accelerator Laboratory}, \orgaddress{\city{Menlo Park}, \state{CA}, \country{USA}}}

\affil[9]{\orgdiv{Chemical Sciences and Engineering Division}, \orgname{Argonne National Laboratory}, \orgaddress{\city{Lemont}, \state{IL}, \country{USA}}}

\abstract{Capturing the structural changes that molecules undergo during chemical reactions in real space and time is a long-standing dream and an essential prerequisite for understanding and ultimately controlling femtochemistry. A key approach to tackle this challenging task is Coulomb explosion imaging, which has benefited decisively from recently emerging high-repetition-rate X-ray free-electron laser sources. With this technique, information on the molecular structure is inferred from the momentum distributions of the ions produced by the rapid Coulomb explosion of molecules. Retrieving molecular structures from these distributions poses a highly non-linear inverse problem that remains unsolved for molecules consisting of more than a few atoms. Here, we address this challenge using a diffusion-based Transformer neural network. We show that the network reconstructs unknown molecular geometries from ion-momentum distributions with a mean absolute error below one Bohr radius, which is half the length of a typical chemical bond.}

\maketitle

\section*{Introduction}
Imaging molecular structure and, in particular, its temporal evolution is fundamental to understanding and steering ultrafast processes, including chemical reactions \cite{zewail_femtochemistry_2000}. Several experimental techniques have been developed during the last decades to study the evolution of molecular structure on picosecond and femtosecond time scales. Relying on a variety of measurement concepts, these techniques probe different aspects of molecular structure and dynamics with different levels of fidelity. Static molecular structures can be captured with the highest position-space resolution using electron microscopy \cite{egerton2005physical}. Time-resolved measurements of evolving molecular geometries often rely on X-rays or high-energy electrons to infer molecular structure from recorded diffraction patterns \cite{centurion_ultrafast_2022, odate_brighter_2023}. Coulomb explosion imaging (CEI), which takes advantage of Coulomb repulsion of nuclei within molecules that are rapidly stripped of their electrons \cite{vager_coulomb_1989, levin_study_1998, herwig_imaging_2013}, is a less mature technique that can also provide time-resolved information if combined with short laser pulses \cite{stapelfeldt_time-resolved_1998}. Since atomic motion typically unfolds on femtosecond time scales (as determined by molecular vibrations), CEI with intense femtosecond laser or X-ray pulses has been exploited for studying molecular structural changes \cite{Ergler_Spatiotemporal_2006, hishikawa_visualizing_2007, jiang_ultrafast_2010, Hansen_fs_torsion_2012, ibrahim_tabletop_2014, liekhus-schmaltz_ultrafast_2015, endo_capturing_2020, jahnke_inner-shell-ionization-induced_2021, jahnke_direct_2025, li_imaging_2025}. These pulses rapidly ionize the target molecules, causing their atomic constituents to repel and fragment as a result of Coulombic forces. The resulting ion-momentum distributions contain information about the initial geometric configuration of the molecule before ionization \cite{pitzer_direct_2013, Pitzer2016, boll_x-ray_2022, li_coulomb_2022, richard_imaging_2025, green_visualizing_2025}.

In all of the imaging methods discussed above, extracting molecular structure from experimental data requires computational algorithms of varying complexity. For diffraction-based imaging techniques, reliable inversion methods are available. In contrast, a corresponding inversion of measured momentum-space data to molecular geometry is not routinely available in the case of CEI. For CEI, geometry retrieval requires solving a highly nonlinear inverse problem, which is extremely challenging when dealing with molecules containing more than 3-4 atoms. In general, inverse problems involve the reconstruction of hidden causal factors from observable data \cite{ongie_deep_2020, zhao_generative_2023}, which are connected by a forward process. If the forward process is trivial to calculate and the noise distribution of this process is known, inverse problems can be solved with the maximum likelihood estimation or the maximum a posteriori approach. Both approaches are typically implemented with an iterative solver, which requires the forward process to be calculated at each step of the iterations. This makes them unfeasible for solving the CEI inverse problem because, in this case, the forward process is driven by the time-dependent many-body interactions governed by quantum mechanics, which is computationally prohibitive to be integrated into an iterative solver. Consequently, direct reconstruction of molecular geometry from CEI has only been demonstrated in a few cases using a classical implementation of the forward process \cite{legare_laser_2005, kunitski_observation_2015}. Most CEI studies \cite{vager_coulomb_1989, hishikawa_visualizing_2007, jiang_ultrafast_2010, pitzer_direct_2013, ibrahim_tabletop_2014, liekhus-schmaltz_ultrafast_2015, Pitzer2016, endo_capturing_2020, boll_x-ray_2022, jahnke_direct_2025, li_imaging_2025, richard_imaging_2025, green_visualizing_2025} have relied on a single-pass simulation of the forward process to compare with experimental measurements, leaving accurate, general reconstruction of molecular geometries an open and unresolved problem.

In this work, we address the molecular structure retrieval problem in CEI with a deep generative neural network designed to reconstruct molecular geometries from ion momentum measurements, which we termed MOLEXA (molecular structure reconstruction from Coulomb explosion imaging). It is built on the Transformer architecture \cite{vaswani_attention_2017} and the diffusion generative modeling framework \cite{sohl-dickstein_deep_2015, ho_denoising_2020, song_score-based_2021, song_denoising_2021, karras_elucidating_2022, xu_geodiff_2022}, with a novel memory mechanism implemented in between the Transformer blocks. The complex forward process of CEI not only renders the classical iterative solvers inapplicable, but also poses a severe challenge for deep learning techniques because it is computationally too demanding to generate adequate data for neural network training. To address the issue of data scarcity, MOLEXA uses a two-stage training approach. Stage 1 trains on a large dataset generated using a computationally inexpensive, approximate forward model, while stage 2 fine-tunes the model on a smaller, high-quality dataset derived from ab initio simulations. The dual-phase strategy reduces the mean absolute prediction error to less than one atomic unit, or half the length of a typical chemical bond. Our present work focuses on the reconstruction of the molecular structure from CEI measurements using X-ray pulses, but the demonstrated generative modeling approach can also be applied to building reconstruction models for CEI measurements using optical lasers \cite{pitzer_direct_2013, bhattacharyya_strong-field-induced_2022, lam_differentiating_2024} and highly charged ion beams \cite{yuan_coulomb_2024}.

\section*{Results}

\subsection*{The MOLEXA network}

\begin{figure}[H] 
	\centering
	\includegraphics[width=0.85\textwidth]{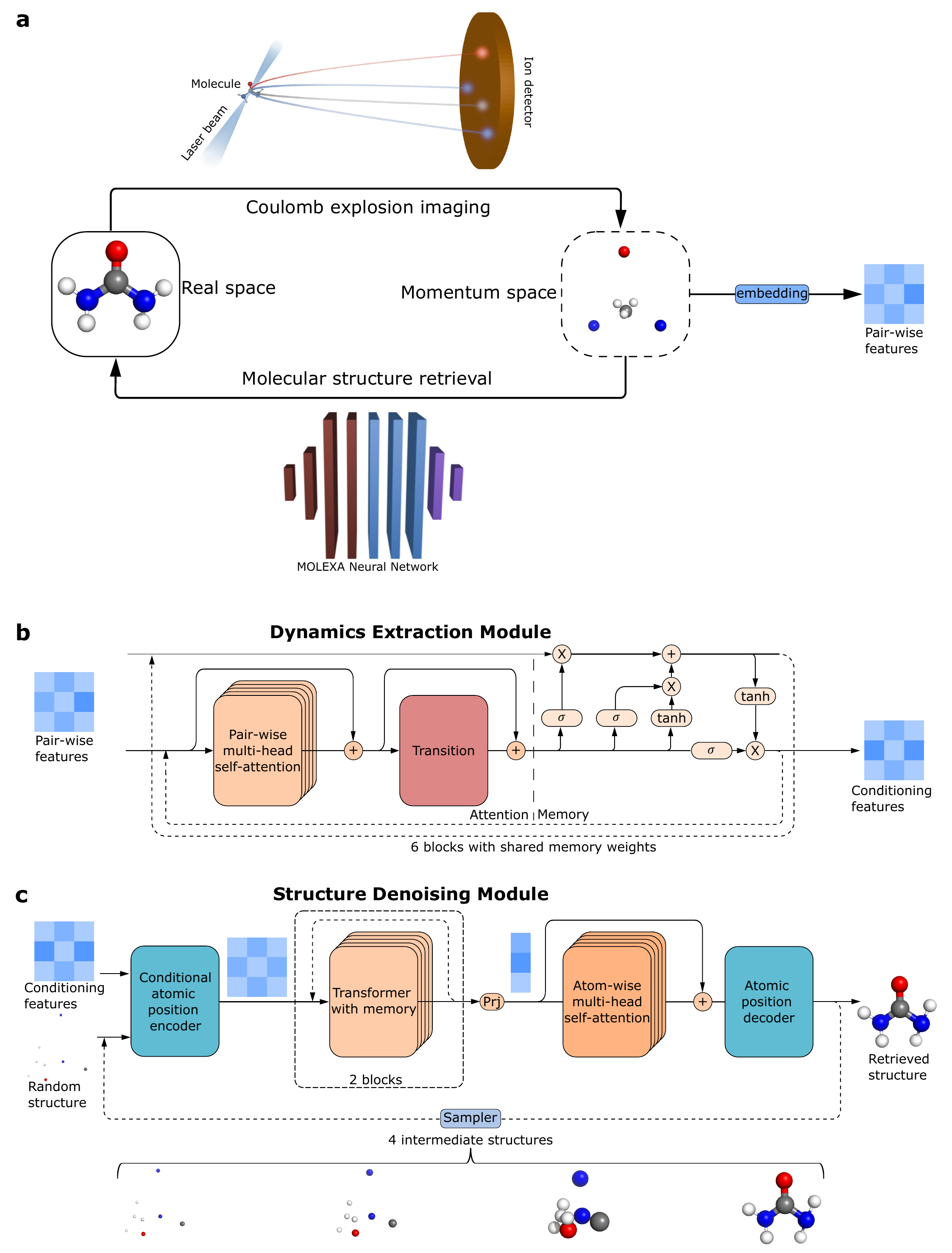} 
	\caption{\textbf{Generative-modeling-enabled molecular structure retrieval from Coulomb explosion imaging.}
		\textbf{a,} Illustration of the Coulomb explosion imaging technique and the molecular structure retrieval from its momentum measurements using the MOLEXA neural network. The main architectural details of MOLEXA are displayed in panels \textbf{b} and \textbf{c}. The ball-and-stick models in this and the subsequent figures represent the scaled spatial arrangement of the atomic constituents in the molecules. \textbf{b,} Dynamics Extraction Module.
        \textbf{c,} Structure Denoising Module.
        }
	\label{fig:illustration_and_model} 
\end{figure}

The MOLEXA model takes the measurable quantities (i.e., the three-dimensional ion momenta measured in coincidence) from CEI as an input and predicts the initial structure of a molecule before its interaction with the X-ray pulse (Fig.~\ref{fig:illustration_and_model}a). It comprises four modules for input embedding, dynamics extraction, structure denoising, and uncertainty estimation, which will be briefly described in the following. Full network details can be found in Supplementary Note 1.

The input to the Embedding Module (Fig.~\ref{fig:illustration_and_model}a) contains the atomic number, charge state, and molecular-frame momentum of each atomic fragment. The embeddings of the atomic number and charge state are concatenated with the linear projection of the momentum to form atom-wise features. The atomic features are concatenated to create pairwise features, which are then processed by a residual block before being sent as input to the Dynamics Extraction Module.

\newpage
The Dynamics Extraction Module (Fig.~\ref{fig:illustration_and_model}b) generates conditioning information used in the Structure Denoising Module. The basic Transformer block, which includes multi-head self-attention, is implemented and accounts for the majority of the computational load for both this and subsequent modules. Instead of directly stacking the Transformer blocks on top of one another, we found that adding memory operations at the end of each block enhances model performance. Similar to the long short-term memory mechanism \cite{Hochreiter-1997-NeuralComp}, the memory operations (on the right side of Fig.~\ref{fig:illustration_and_model}b) include a forget gate that regulates what information to discard from the previous state, an update gate that decides which information in the Transformer output should be added to the memory, and an output gate that selectively sends the current state of the memory to the next Transformer block. In comparison to using skip connections, the memory mechanism is found to suppress the mean (maximum) atomic distance and angle errors by 3.6\% (4\%) and 1.3\% (4.4\%), respectively. We refer to the combination of the Transformer and memory operations as the "Transformer with Memory" (TM) block. There are six TM blocks in the Dynamics Extraction Module.

The Structure Denoising Module, illustrated in Fig.~\ref{fig:illustration_and_model}c, reconstructs the molecular structure using a reverse diffusion process. It starts with a noisy molecular structure. Its atomic positions are encoded on the basis of the output of the Dynamics Extraction Module and the current noise level. Pairwise features derived from this encoding are processed by two TM blocks. The output is projected to obtain atom-wise features that are further processed through a self-attention block. The Position Decoder takes the transformed atomic features and outputs a less noisy molecular structure. During inference, this structure is iteratively refined by a diffusion sampler with a noise schedule adapted from Ref. \cite{karras_elucidating_2022}. As shown in Fig.~\ref{fig:illustration_and_model}c, five iterations, corresponding to four intermediate structures, are performed to obtain the final molecular structure. The smaller atom sizes shown for the earlier iterations reflect larger interatomic distances.

The Uncertainty Estimation Module is trained to match the predicted uncertainty with the absolute error between the predicted and ground-truth structures. It can provide uncertainty estimations for the structure predictions. Using pairwise features from the Dynamics Extraction Module and the predicted molecular structure, the Uncertainty Estimation Module estimates the errors of the predicted atomic positions using two TM blocks, followed by an uncertainty decoder. It pre-defines the uncertainty bins \( \mathbf{r} = [0, 0.05, 0.1, \dots, 9.95] \) and estimates the probability that the prediction error falls within each of these bins. For each coordinate of the predicted atomic positions, the uncertainty is then calculated as the probability-weighted sum of the bin values.

\subsection*{Training}
\label{subsec_training}

Unlike text or image generation models, for which there exists an enormous amount of training data, deep learning models in physical sciences often face the data scarcity issue, which is one of the main obstacles preventing the widespread adoption of deep learning techniques for solving physics-related problems. For the molecular structure retrieval problem, we created two training datasets by performing Coulomb explosion simulations at two levels of theory. One level involves the ab initio calculation of the XFEL-induced Coulomb explosion of molecules, which tracks the quantum transition probabilities across all participating electronic configurations while treating the nuclei as moving in a classical force field \cite{Ho-2017-JPB}. It has been shown to produce results that agree with experiments \cite{li_imaging_2025}. A similar level of theory was used, for example, in Refs. \cite{boll_x-ray_2022, richard_imaging_2025}. These high-level simulations are computationally expensive. Thus we created only a small dataset containing 76,000 samples, using a thousand CPUs for more than a month. A portion of this dataset was kept for validation (10\%) and testing (10\%) purposes. Since the computation time scales roughly exponentially with the number of atoms, these simulations were limited to molecules with fewer than ten atoms, which was a compromise to balance the dataset size with computational constraints. The second level of theory is a much cheaper, classical Coulomb explosion model with crude approximations \cite{li_coulomb_2022}. It was used to generate a dataset that is about a hundred times larger. MOLEXA was first trained on this large but inaccurate dataset and then on the small dataset, which is more accurate and best reflects the reality of a CEI experiment. We found that the two-stage training approach reduced the structure prediction error by a factor of two compared to training solely on the smaller but more accurate dataset.

Before training in each stage, both the ion-momentum and ground-truth position distributions were centered and aligned to a common molecular frame. Specifically, for each molecule, the emission direction of the heaviest ion fragment is defined as the \textit{x} axis. Then the cosine similarities of the momenta of the other ion fragments relative to the \textit{x} direction are calculated. The ion momentum with the smallest cosine similarity is chosen to define the \textit{y} axis through Gram-Schmidt orthonormalization so that this momentum vector would fall within the first quadrant of the \textit{x-y} plane. Both the momenta of all ion fragments and the initial molecular structure, after being centered at the origin, are transformed into this coordinate frame. For diatomics, the coordinate system is fully determined once the emission direction of the heavier ion fragment is defined as the \textit{x} axis. This pre-alignment procedure, applied consistently across all molecules, transforms the structure retrieval task from one with arbitrary coordinate frames to one within a fixed molecular frame, thereby eliminating the need to explicitly incorporate translational and rotational invariance into the model.

The loss function consists of two parts: a weighted mean squared error for the predicted molecular structures, and a cross-entropy loss for uncertainty estimations. Both parts were used throughout the two training stages. The Uncertainty Estimation Module was further fine-tuned using the validation dataset while keeping the other modules frozen, during which only the second part of the loss function was used. All reported results were generated from the test dataset. The training and validation datasets only contain molecules with fewer than eight atoms, while the eight- or nine-atom molecules were set aside to test the generalization capability of the model. Additional details on loss function, training, and testing are provided in the Methods section.

\subsection*{Model Performance}

Using the test dataset of molecules with less than eight atoms, the mean absolute error (MAE) is 0.52~a.u. (atomic units), the mean (maximum) distance error (DE) of all atomic pairs in a molecule is 0.98 (2.11) a.u., and the mean (maximum) angle error (AE) of all directional triplets in a molecule is 13.97 (38.39) degrees. MOLEXA, which was trained on only molecules with up to seven atoms, is also capable of reconstructing the structure of molecules containing eight or nine atoms. For these molecules larger than those included in the training dataset, the MAE is 0.66~a.u., the mean (maximum) DE is 1.16 (3.43)~a.u., and the mean (maximum) AE is 14.12 (57.91) degrees. The inference time distribution evaluated on all test molecules is plotted in Supplementary Fig.~\ref{fig:S_inference_time} and has a mean of 59.8 ms. Each of the reconstructed molecular structures is from a single model prediction. Based on the standard deviations of the model predictions plotted in Supplementary Fig.~\ref{fig:S_uncertainty_std}, the expected discrepancy between the corresponding atomic coordinates of two independent reconstructions for a single molecule is about 0.07~a.u.

\begin{figure} 
	\centering
	\includegraphics[width=1\textwidth]{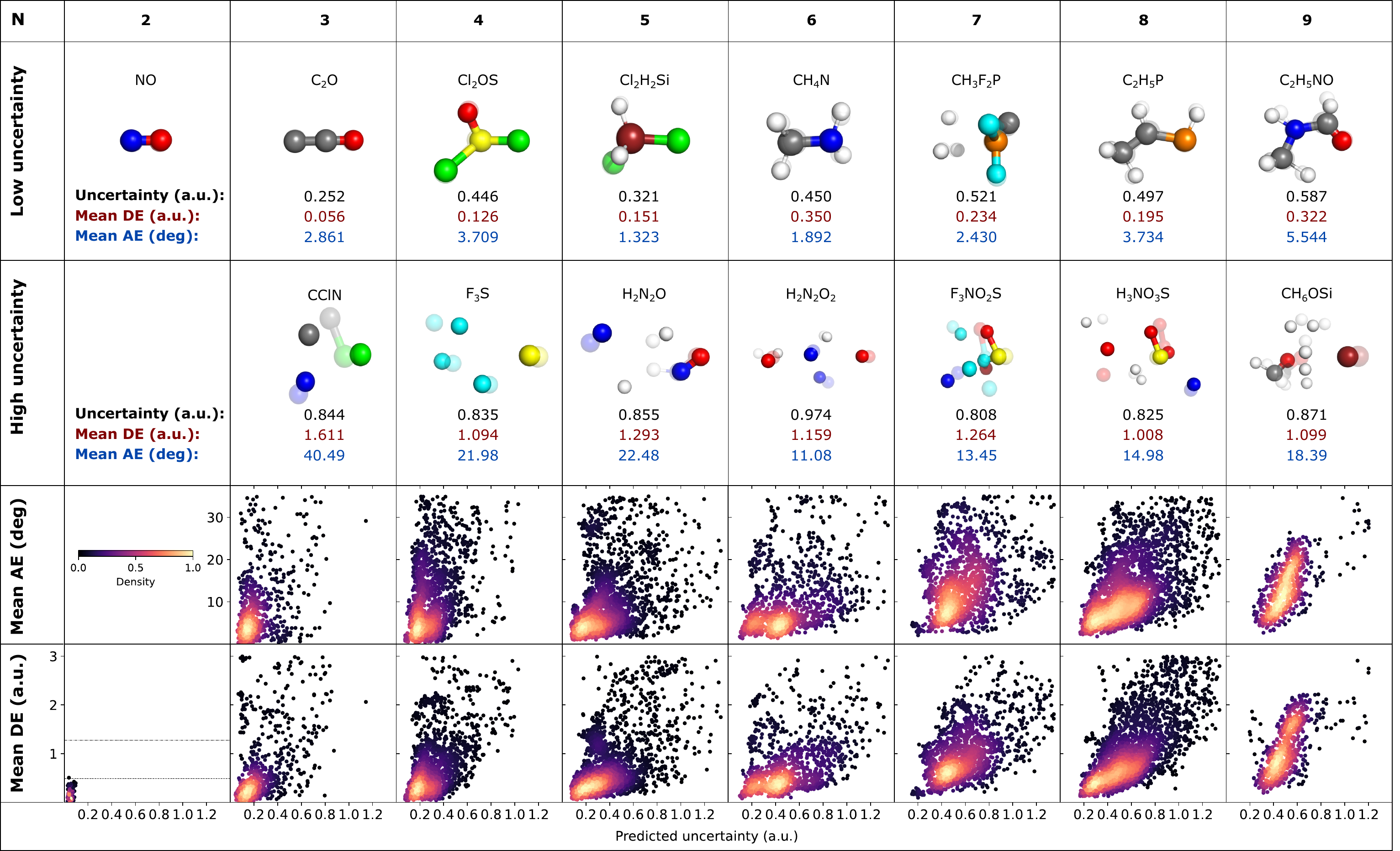} 
	\caption{\textbf{Overview of MOLEXA's reconstruction performance and its relation to the predicted uncertainties.} The columns from left to right represent molecules with an increasing number of atoms. Top row: Exemplary structure predictions with low predicted uncertainties. The predicted and ground-truth structures are plotted as opaque and semi-transparent ball-and-stick models, respectively. The corresponding uncertainty, mean DE, and AE are listed below each molecular structure. The color coding of the elements is as follows - H: white, C: gray, N: blue, O: red, F: cyan, Si: brown, P: orange, S: yellow, and Cl: green. The ball-and-stick models were plotted with PyMOL which considers two atoms bonded if their distance is smaller than a tolerance-expanded sum of their covalent sizes. Second row: Exemplary structure predictions with high predicted uncertainties. The maximum DE and AE values for these molecules are summarized in Supplementary Tables~\ref{tab:max_de_ae_low}-\ref{tab:max_de_ae_high}. The two bottom rows depict the dependence of the mean DE and AE on the predicted uncertainty. The density plots use the same color scale. The dash-dotted line in the bottom left corner marks the mean DE (1.27~a.u.) of reconstructions with the classical $\frac{1}{KER}$ model, and the dashed line marks the error (0.49~a.u.) for the optimized empirical model $\frac{0.797}{KER}$. The corresponding MOLEXA input data including the ion charge states and momentum distributions are shown in Supplementary Fig.~\ref{fig:S_momen_performance}.}
	\label{fig:performance} 
\end{figure}

Figure~\ref{fig:performance} provides an overview of the structure retrieval performance of the model. In each column of the figure the results for molecules consisting of $N$ atoms are presented, showing exemplary structures predicted with low (top row) and high (second row) reconstruction uncertainties. The two bottom rows indicate how the general performance of MOLEXA (in terms of mean DE and AE) behaves as a function of the predicted uncertainty for molecules of different sizes. The Uncertainty Estimation Module produces uncertainties for all coordinates of all atoms in a molecule. The overall uncertainty displayed here was obtained by taking an average of the uncertainties of all atomic positions of a molecule. The heatmaps in Fig.~\ref{fig:performance}, as well as those in Supplementary Figs.~\ref{fig:S_accuracy_mae_uncertainty}-\ref{fig:S_max_DE_AE} for MAE, maximum DE and AE, show that there is a strong correlation between the predicted uncertainty and the errors of the reconstructed molecular structures. This indicates that the former can serve as a reliable metric for assessing whether a MOLEXA reconstruction is trustworthy or not. It can be observed from these heatmaps that the predicted uncertainties can underestimate the reconstruction errors, especially for larger molecules with larger prediction errors. We attribute this to a training bias caused by the Uncertainty Estimation Module being trained on reconstructions most of which have relatively low errors. Such bias can potentially be mitigated with a weighted cross-entropy loss in future developments.

With MOLEXA being a diffusion model, an uncertainty quantification for each sample is also obtainable by calculating the standard deviation of an ensemble of its structure predictions. The uncertainties calculated with these two approaches are shown in Supplementary Fig.~\ref{fig:S_uncertainty_std} to have an approximately linear relationship. The former method is used here because it provides an uncertainty estimate for each prediction without requiring an ensemble of predictions. Note that although an uncertainty estimate can assist the assessment of the trustworthiness of a prediction, it is not effective in selecting the most accurate reconstruction out of many predictions for a molecule. This is indicated by the small correlation coefficients between uncertainty estimate and prediction errors plotted in Supplementary Fig.~\ref{fig:S_err_uncertainty_corr}.

Figure~\ref{fig:performance} also depicts the dependency of prediction errors on the size of the molecules examined. For diatomics, the mean DE is 0.155 a.u. As a reference, the mean DE's from calculating the bond length of the same set of diatomic molecules with the classical $\frac{1}{KER}$ model and the optimized empirical $\frac{0.797}{KER}$ model are 1.27 and 0.49 a.u., respectively. (KER stands for the kinetic energy release from Coulomb explosion. The empirical factor 0.797 was determined by minimizing the error of the $\frac{c}{KER}$ model on the test data.) For larger molecules, the average DE and AE distributions gradually shift upwards, implying that it becomes more difficult for the model to learn the underlying X-ray-induced dynamics in molecules containing more atoms. Although the increased prediction error for molecules with eight or nine atoms is still acceptable, it is expected to increase even more for larger molecules. Corresponding example predictions for the 1,3-cyclohexadiene molecule containing fourteen atoms are shown in Supplementary Fig.~\ref{fig:S_CHD}. As expected, they show large discrepancies from the ground-truth structures. In addition to the dependence on the number of atoms, predictions for molecules with larger atomic distances tend to result in larger errors as indicated by Supplementary Fig.~\ref{fig:S_DE_AE_distance}. Training with a more diverse dataset that includes larger molecules with larger atomic distances would be needed to break these limitations and further extend the applicability of MOLEXA in the future. 

The model has two other major limitations. One originates from the non-unique mapping between ion momentum vectors and real-space molecular geometries as revealed by Ref. \cite{sayler_nonunique_2018}, which can increase the overall prediction error.  In future developments, the non-uniqueness issue can be mitigated by a model that makes use of multiple coincidence channels as input data, in contrast to a single channel used by the current model. The other limitation is that the input data must include all ions produced by Coulomb explosion of the molecule. But in a CEI experiment, the complete coincident ion detection from each explosion is a relatively low-probability event, especially for larger systems. Our current way to circumvent this limitation is by distilling the required full-coincidence momentum vectors from the data accumulated over many explosion events \cite{boll_x-ray_2022, li_coulomb_2022, richard_imaging_2025, green_visualizing_2025}. This approach is used to prepare the input data of the experimentally studied molecules discussed in the next section. Further development of the neural network is needed to make it capable of molecular structure prediction from partial coincidence sets, which would be particularly beneficial for single-event reconstruction.

\subsection*{Application of MOLEXA}

In this subsection, we first demonstrate the ability of MOLEXA to perform the inversion of experimental data into real space molecular geometries. For this, we used the data acquired during several experiments carried out at the European X-ray Free-Electron Laser facility to reconstruct the equilibrium geometry of ground-state molecules including water, tetrafluoromethane, and ethanol. No further molecule-specific input was provided to the model for retrieving the structures. The experiments were performed in multiple CEI beamtimes using the COLTRIMS (Cold Target Recoil Ion Momentum Spectroscopy) Reaction Microscope \cite{Dorner2000} at the Small Quantum Systems (SQS) instrument. The beamtimes and X-ray pulse parameters are summarized in Supplementary Table~\ref{tab:exp}. In all experiments, the molecular samples were delivered to the interaction region through a supersonic expansion followed by three skimmers and an adjustable collimator. The distance from the nozzle to the interaction region is about 54~cm. The pressure in the main chamber was maintained at 1~$\times$~10$^{-11}$~mbar. The focal spot size of the X-ray beam was about 1.5~--3~$\mu$m and the X-ray pulse duration was less than 25~fs based on the 250pC electron bunch charge. The ion fragments produced in the interaction region were guided by a homogeneous electric field to a time- and position-sensitive detector. The lab-frame momentum vectors of the ion fragments were then reconstructed from the detector readouts, which were subsequently transformed into the molecular frame with the alignment procedure discussed in the Training subsection.

In real molecules, the nuclear ground state exhibits a spatial distribution due to the uncertainty principle, such that different nuclear geometries are sampled even in the absence of excitation. Because the mapping between ion momentum and molecular geometry is nonlinear, averaging in momentum space does not strictly correspond to averaging in real space. In principle, the equilibrium molecular structure is best estimated by reconstructing geometries from individual single-shot, full-coincidence Coulomb explosion events and averaging the predicted structures in real space. But single-shot full-coincidence detection remains challenging for today’s CEI experiments. In the following, the structures are instead predicted from the coincident momentum vectors (Supplementary Figs.~\ref{fig:S_H2O} - \ref{fig:S_C2H6O_3}) obtained by averaging over the single-shot data. This procedure yields an approximate equilibrium geometry from the ensemble-averaged observables, rather than the full nuclear probability distribution or any single instantaneous molecular configuration.

\begin{figure}[H] 
	\centering
	\includegraphics[width=1\textwidth]{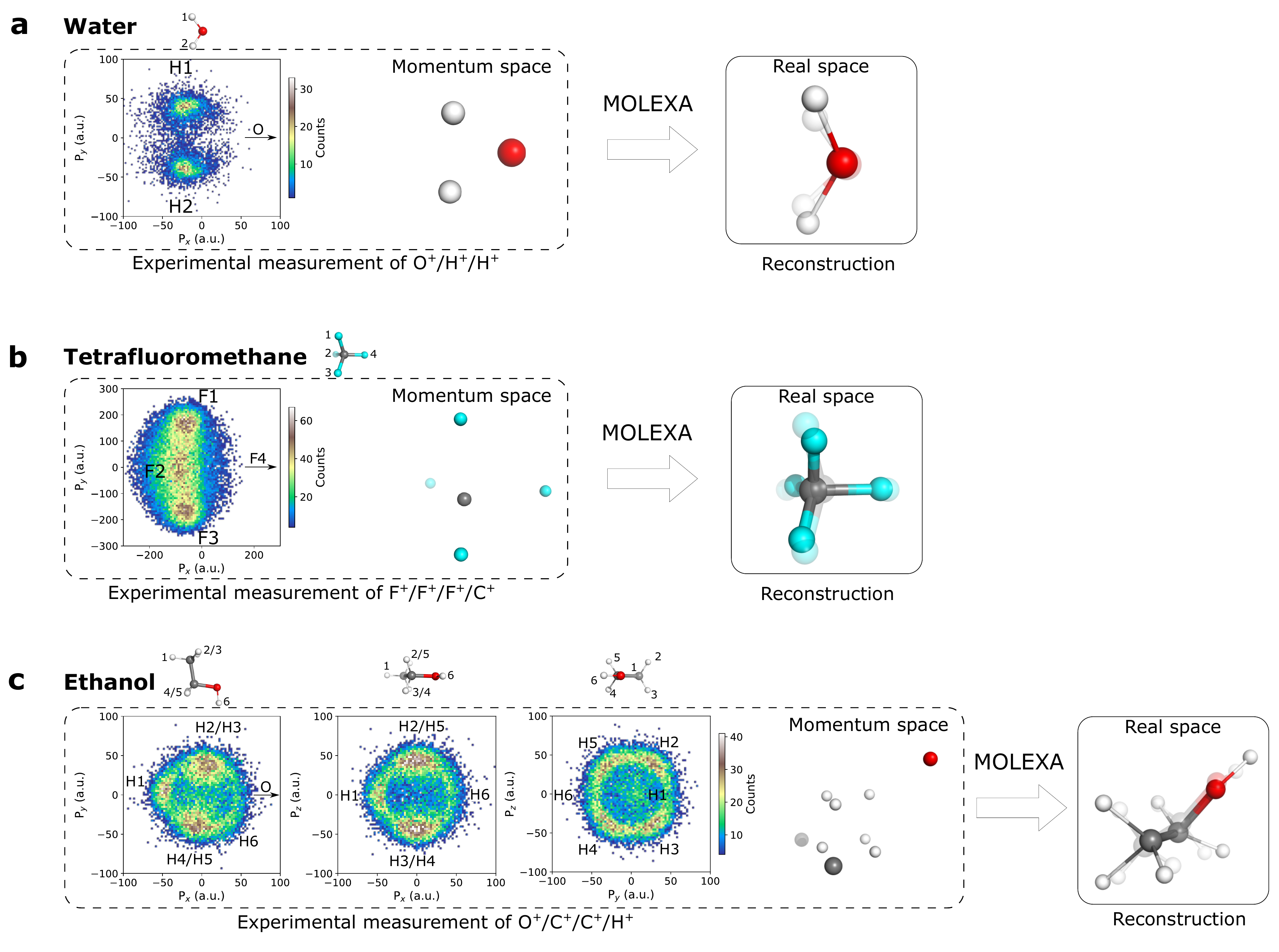} 
	\caption{\textbf{Reconstruction of molecular geometries from experimental data.}
		\textbf{a,} Molecular structure reconstruction of water. The measured 2D momentum map is shown on the left. To its right is the illustration of the averaged momentum distribution of the three ion fragments, which is used as the input to MOLEXA. In the real space, the reconstructed and ground-truth structures are plotted as opaque and semi-transparent ball-and-stick models, respectively. \textbf{b,} Molecular structure reconstruction of 
 tetrafluoromethane. \textbf{c,} Molecular structure reconstruction of ethanol. The corresponding orientations of the pre-explosion molecule are displayed at the top of the 2D momentum maps. The color coding of the elements is as follows - H: white, C: gray, O: red, and F: cyan. The ground-truth structures are from the NIST Computational Chemistry Comparison and Benchmark Database \cite{CCCBDB_2022_release22}.}
	\label{fig:reconstruction} 
\end{figure}

As a first example, Figure~\ref{fig:reconstruction}a shows the reconstruction of the molecular structure of water molecules. The employed dataset \cite{water_beamtime} used in this analysis is identical to the one used in Ref. \cite{jahnke_inner-shell-ionization-induced_2021}. The 2D heatmap in the left-most part of the panel displays the experimentally measured molecular-frame momentum distribution of two protons detected in coincidence with a singly charged oxygen ion. Next to it in the middle, we show an illustration of the centroids of the momentum distributions of the three ions, which serve as the input for MOLEXA. The results obtained from the reconstruction are shown on the right. The reconstructed molecular geometry (opaque) is plotted on top of the ground truth (semi-transparent), with the corresponding MAE being 0.296~a.u. The mean (maximum) DE and AE are 0.674 (1.199)~a.u. and 18.459 (27.689)~degrees, respectively. Next, we test the model on tetrafluoromethane, a molecule consisting of five atoms. The corresponding reconstruction is shown in Fig.~\ref{fig:reconstruction}b. The left-most panel depicts the momenta of the three of the four fluorine ions in a molecular frame spanned by the fourth fluorine ion and one of the three. The reconstructed position-space geometry has a MAE of 0.238~a.u., a mean (maximum) DE of 0.66 (1.117)~a.u., and a mean (maximum) AE of 5.943 (17.173)~degrees. The data was recorded during the commissioning of the SQS reaction microscope \cite{tetrafluoromethane_beamtime}.  

As a benchmark for an application to molecules with up to nine atoms, we applied MOLEXA to a CEI dataset recorded for ethanol molecules \cite{ethanol_beamtime}. The results are shown in Fig.~\ref{fig:reconstruction}c. The 2D maps show the molecular-frame momentum distributions of protons in the coincidence channel O$^+$/C$^+$/C$^+$/H$^+$, viewed from three different perspectives. We added the corresponding orientation of the real-space molecule to the top of each graph to aid the identification of the six protons in the momentum maps. The input to the model is again obtained by taking the centroids of the momentum distributions of the nine ions. The retrieved molecular geometry plotted together with the ground truth at the right has a MAE of 0.429~a.u. The mean (maximum) DE and AE are 1.024 (2.007)~a.u. and 9.011 (32.319)~degrees, respectively. More details on the reconstruction from experimental data, including the momentum distributions of other ions not displayed in Fig.~\ref{fig:reconstruction}, the momentum centroid data, and the reconstructed atomic coordinates as well as the predicted error estimates, can be found in Supplementary Figs.~\ref{fig:S_H2O}-\ref{fig:S_C2H6O_3} and Supplementary Tables~\ref{tab:h2o}-\ref{tab:c2h6o}.

\begin{figure}[H] 
	\centering
	\includegraphics[width=1\textwidth]{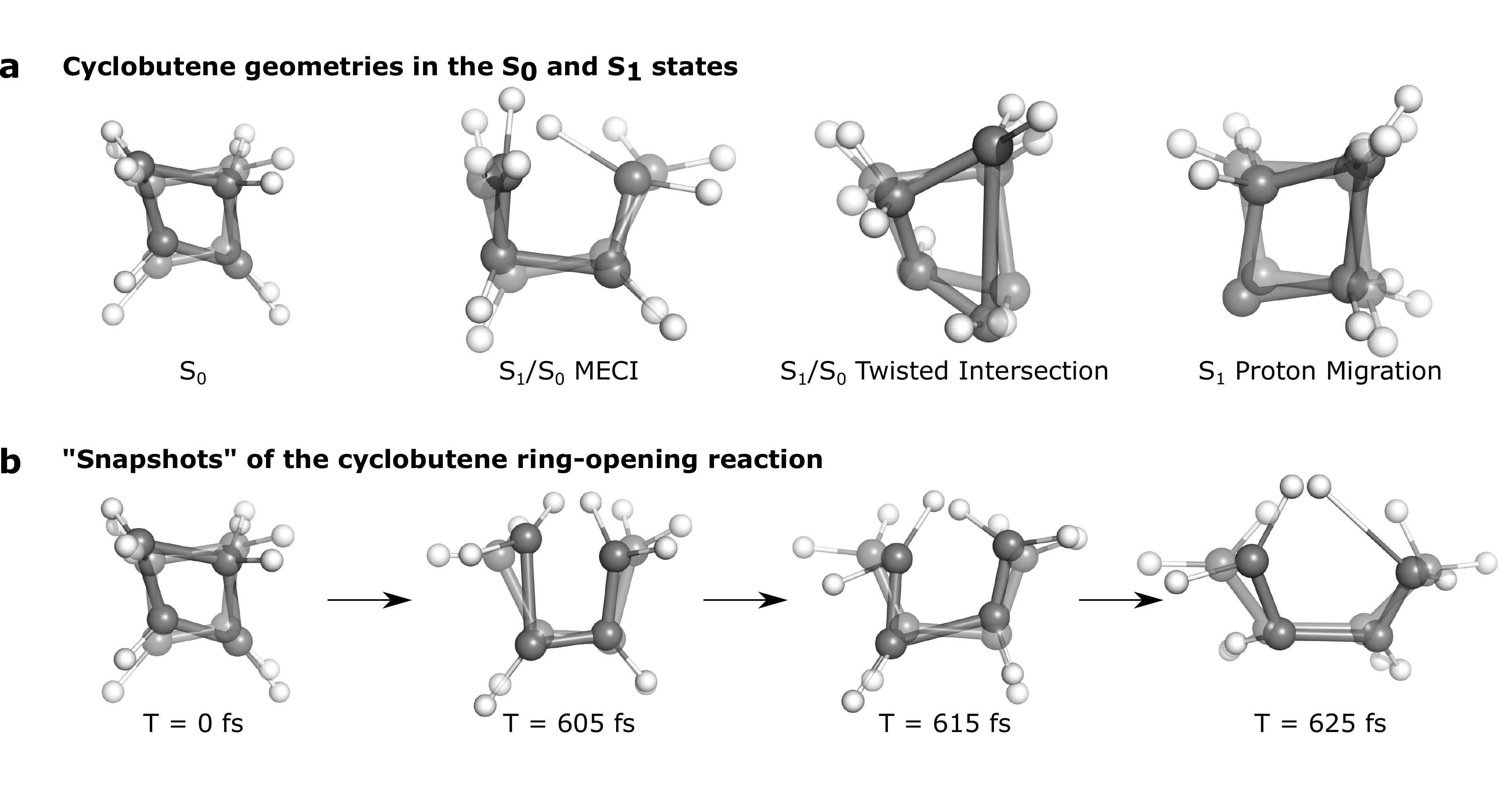} 

	\caption{\textbf{Reconstruction of structural changes.}
		\textbf{a,} Reconstructed geometries of cyclobutene in its S$_0$ and S$_1$ states. \textbf{b,} Reconstructed "snapshots" of cyclobutene during a chemical reaction. The color coding of the elements is as follows - H: white and C: gray.}
	\label{fig:reaction} 
\end{figure}

The ultimate aim of CEI is to directly observe molecular dynamics during a chemical reaction in a time-resolved manner. In order to achieve this, coincident momentum-space fragmentation patterns are measured at different instants during the chemical reaction, thus allowing to study the molecular structural changes as the chemical reaction unfolds on femtosecond or longer time scales. In the following example, we exploit MOLEXA to reconstruct the different geometries of cyclobutene as predicted by ab initio simulations \cite{cyclobutane_ong_2010}. The electrocyclic reactions of cyclobutene represent a textbook example of pericyclic reactions that are among the important classes of chemical reactions in organic chemistry. Figure~\ref{fig:reaction}a shows that MOLEXA is capable of reconstructing different possible geometrical changes, including ring opening, twisting, and proton migration, after cyclobutene is excited from the ground state (S$_0$) to the S$_1$ state. In Fig.~\ref{fig:reaction}b, MOLEXA is used to reconstruct position-space "snapshots" of cyclobutene as it undergoes a ring-opening reaction. The reconstructions show that the model can provide insight about the overall structure and identify gross structural rearrangements such as proton migration and ring opening. Further details on the reconstruction of cyclobutene geometries can be found in Supplementary Tables~\ref{tab:c4h6_s0}-\ref{tab:c4h6_t_625}. More examples demonstrating the model's capability to reconstruct varying structures of molecules are displayed in Supplementary Fig.~\ref{fig:S_multi_structures}. It is worth noting that the reconstructions in Fig.~\ref{fig:reaction} are idealized scenarios where the input momentum vectors correspond to a single well-defined molecular geometry. This can work if the input coincidence data is assembled from a single explosion event. For time-resolved CEI experiments, the accumulated data for each time delay typically results from a mixture of geometries corresponding to multiple quantum states. Classification of the experimental data into individual states, based on, e.g., ion charge-state characteristics and kinetic energies \cite{li_imaging_2025}, is hence required prior to applying MOLEXA for reconstruction.

\section*{Discussion}

MOLEXA is a powerful neural network designed for molecular structure reconstruction with the CEI technique. It allows for inverting momentum-space datasets to position space, providing the structure of a molecule right before its explosion by an X-ray pulse and showing particular effectiveness in reconstructing the overall structure of molecules. In addition,  it is capable of providing an uncertainty estimate for its reconstructed molecular geometries. By employing time-resolved CEI datasets, MOLEXA has the potential to provide "snapshots" of a molecule at different instants during a chemical reaction, which can enable the use of the CEI technique for direct reconstruction of molecular dynamics in position space as they unfold on their natural time scales.

Apart from taking advantage of recent advances in deep learning, such as the Transformer framework and diffusion-based generative modeling, MOLEXA utilized the "Transformer with Memory" architecture and went through a two-stage training, both of which were crucial for achieving its effectiveness in molecular structure predictions. It demonstrates the potential of generative modeling in solving inverse problems that classical approaches cannot address due to the excessive complexity of their forward models, which prevents integration into an iterative procedure. Even with deep learning techniques, solving such problems poses a challenge because of training data scarcity. The two-stage modeling can be applied as a general approach to addressing this issue when a complicated forward process can be approximated as a simple model.

\bibliography{sn-article}

\section*{Methods}

\subsection*{Dataset creation}
For the ab initio simulation, we used a theoretical model that combines Monte Carlo / Molecular Dynamics simulations (MC/MD) \cite{Ho-2017-JPB} with a classical over-the-barrier (COB) model \cite{ryufuku_oscillatory_1980, niehaus_classical_1986, HoJCP2023} to track inner-shell photoionization, Auger-Meitner cascades, valence electron redistribution, and nuclear dynamics. Photoabsorption and inner-shell cascade processes were modeled using a Monte Carlo method to calculate quantum electron transition probabilities across all participating electronic configurations (ECs), including ground, core-excited, and valence-excited states. The electronic-structure calculations were based on the relativistic Hartree-Fock-Slater (HFS) method, which provided bound-state and continuum wavefunctions for computing cross sections of photoionization, shake-off, electron-impact ionization, and electron-ion recombination, as well as Auger-Meitner and fluorescence decay rates. The molecular-dynamics component tracked the motion of atoms, ions, and delocalized ionized electrons. The COB model simulates electron-transfer dynamics in the valence shell. In this model, an electron fills a vacancy in the valence shell of a neighboring atom when its binding energy is higher than the Coulomb barrier. When the atoms are far apart, the resulting Coulomb barrier suppresses electron transfer. Electron transfer takes place instantaneously when the electron orbital energy is higher than the Coulomb barrier.

With this ab initio model, the Coulomb explosion of three hundred different molecules with fewer than ten atoms was first simulated in their equilibrium geometry. The X-ray pulses have a photon energy of 2~keV, pulse energy of 1~mJ, pulse duration of 15~fs, and focal spot size of 1~$\mu$m. In order to expand the dataset, the simulations were additionally performed tens of times on each of these molecules after randomly varying their structures. Because of the stochastic nature of the X-ray interaction with molecules, the atomic charge-state combination of the resulting ion fragments from a molecule can vary from one simulation trajectory to another. With a hundred thousand trajectories simulated for each molecule at a fixed structure, the number of trajectories ending at each of the possible charge-state combinations was enumerated. Only combinations with a count greater than three hundred were considered. The momentum of each ion fragment was obtained by averaging all trajectories. For every such charge-state combination, the atomic number, charge state, and momentum of all ion fragments, as well as the initial atomic coordinates of the molecule, were included into the dataset as a single entry. With an average of ten structures simulated for each of the three hundred molecules and an average of about ten charge-state combinations produced from a molecule at a particular structure, the dataset contains 76~000 entries. It was further split into training (80\%), validation (10\%), and test (10\%) datasets. The training dataset was used in the second step of the two-stage modeling process. Exemplary samples from the test dataset are shown in Supplementary Fig.~\ref{fig:S_samples_predictions} together with the corresponding predictions.

Because the ab initio simulation was computationally expensive and could only be used to generate a small dataset, an approximate forward Coulomb explosion model \cite{li_coulomb_2022} was used to create a dataset about two orders of magnitude larger. The model describes the charge-up of each atom in a molecule with a modified error function that increases from zero to the final charge number within a time window controlled by the constant $\tau$. For the simulation, $\tau$ was set to be 45~fs, which was determined by minimizing the discrepancy of the results of the approximate model with respect to those of the ab initio simulations. Using the time-dependent charge states given by the modified error function, the Coulomb explosion dynamics were simulated with the Runge-Kutta approach that propagates the time-dependent positions and velocities of ion fragments according to classical mechanics. The "molecules" used for this approximate simulation were generated by enumerating all possible combinations of the 9 elements (H, C, N, O, F, Si, P, S, and Cl), with the number of atoms in each combination less than 10. The positions of the atoms in a "molecule" were sampled from a uniform distribution ranging from -10~a.u. to 10~a.u. The dataset produced from the simulation with these "molecules" consists of six million entries and was used for the first step of the two-stage modeling process.

\subsection*{Training details}
During training, the reverse diffusion process (Structure Denoising Module) was run only once for each training step. Instead of taking a random structure as input, it starts with a noisified ground-truth structure with the noise level controlled by \( \sigma_i \). The Structure Denoising Module was trained to denoise this input and generate a geometry \( \mathbf{G}_i^{prediction} \) that is a reconstruction of the ground truth \( \mathbf{G}_i^{ground\_truth} \). The corresponding loss function is

\begin{equation}
\mathcal{L}_{\text{x}} = \mathbb{E}_i \left( w_i \| \mathbf{G}_i^{prediction} - \mathbf{G}_i^{ground\_truth} \|_2^2 \right),
\end{equation}

where the weight \( w_i \) is set according to Ref.~\cite{karras_elucidating_2022} and given by

\begin{equation}
w_i = \frac{\sigma_i^2 + \sigma_{data}^2}{\sigma_i^2\sigma_{data}^2},
\end{equation}

with $\sigma_{data}$ determined by the standard deviation of the molecular structures in the dataset. In addition to structure reconstruction, MOLEXA was trained to estimate the uncertainty of its predicted structures. As already mentioned in the main text, it first gets the probability \textit{s$_n^i$} that the uncertainty of the \( i^{th} \) predicted coordinate \( x_i^{prediction} \) falls into the \( n^{th} \) bin of the pre-defined uncertainty list \( [r_0, \dots, r_{200}] \). The absolute error \( |x_i^{prediction} - x_i^{ground\_truth}| \) is classified according to this list as a one-hot encoded vector $\mathbf{q}^i$. The loss function for the uncertainty estimate is then calculated as the averaged cross entropy

\begin{equation}
\mathcal{L}_{\text{u}} = -\mathbb{E}_{n,i} \left(q_n^ilog(s_n^i) \right).
\end{equation}

The combined loss function used during training is given by

\begin{equation}
\mathcal{L} = c_x\mathcal{L}_{\text{x}} + c_u\mathcal{L}_{\text{u}}, 
\end{equation}

where \( c_x \) and \( c_u \) are the weights of the structure retrieval and uncertainty estimation loss functions, respectively.

 The neural network was trained through two stages. The weights were initialized using the orthogonal Glorot initialization \cite{glorot_understanding_2010, saxe_exact_2014} with a scale of 2 for the linear layers and sampled from the uniform distribution with a range from $-\sqrt{3}$ to $\sqrt{3}$ for the embedding layers. In the first stage, it was trained on the large dataset generated by the approximate forward model. The weight \( c_x \) in the loss function was set to 400. And \( c_u \) was set to 0.1 for the first seven epochs and 1 afterwards. For the second stage, the training was performed with the dataset which is about a hundred times smaller and generated by the ab initio forward model. The weights \( c_x \) and \( c_u \) in the loss function were set to 400 and 0.01, respectively. In order to improve the accuracy of uncertainty predictions, after the two-stage training, the network was further trained on the validation dataset. Only the Uncertainty Estimation Module was trained while the other modules were kept frozen. The weight \( c_x \) was set to 0 and \( c_u \) to 0.01. During all training phases, the Adam optimizer \cite{kingma_adam_2015} was used for optimization. Its parameters \( \beta_1 \), \( \beta_2 \), and \( \epsilon \) were fixed at 0.9, 0.99, and 10$^{-5}$, respectively. The learning rate was kept at 0.001. Training with learning rate decay was tested, but did not improve the prediction errors. More details on the two-stage training are summarized in Supplementary Table~\ref{tab:training}.

\subsection*{Coordinate frame transformations for experimentally studied molecules}

The molecular frames used by the 2D maps in Fig.~\ref{fig:reconstruction} were defined with a procedure similar to that described in the main text. The flying direction of a reference ion (O$^+$ for water and ethanol, and F$^+$ for tetrafluoromethane, as indicated by the arrows in the 2D maps of Fig.~\ref{fig:reconstruction}) was set as \textit{x} axis. The \textit{y} axis was then defined such that the momentum vector of a second reference ion (H$^+$ for water, and F$^+$ for tetrafluoromethane) falls within the positive \textit{x-z} plane. For ethanol, the second reference ion was chosen to be the C$^+$ ion that has a smaller cosine similarity with respect to the first reference ion (O$^+$). The \textit{y} axis was defined such that the momentum vector of this second reference ion falls within the positive \textit{x-y} plane. The molecular-frame momentum distributions and their centroids are shown for the four molecules in Supplementary Figs.~\ref{fig:S_H2O} - \ref{fig:S_C2H6O_3}. The atomic coordinates reconstructed by MOLEXA together with the ground truth are listed in Supplementary Tables~\ref{tab:h2o} - \ref{tab:c2h6o}.

\section*{Acknowledgements}
We acknowledge the LCLS data team and the SLAC Shared Science Data Facility (S3DF) for providing the compute and data storage used in model development. We acknowledge the teams of the three European XFEL experiments (2150, 2181, and 2926) for sharing the associated data. X.L. would like to thank Patricia Vindel Zandbergen for her help related to the model testing, and Philipp Schmidt for his support in experimental data processing. This work is supported by the Linac Coherent Light Source, SLAC National Accelerator Laboratory, which is funded by the U.S. Department of Energy, Office of Science, Office of Basic Energy Sciences under Contract No. DE-AC02-76SF00515. P.J.H. is supported by the U.S. DOE BES Chemical Sciences, Geosciences, and Biosciences Division under Contract No. DE-AC02-06CH11357. D.R. and A.R. are supported by grant no. DE-FG02-86ER13491 from the same funding agency and also acknowledge dedicated support for ML/AI developments through the GRIPex program at Kansas State University. F.T. acknowledges funding by the Deutsche Forschungsgemeinschaft (DFG, German Research Foundation) -- Project 509471550, Emmy Noether Programme. T.J.A.W. was supported by the Atomic, Molecular, and Optical Sciences Program of the U.S. Department of Energy, Office of Science, Office of Basic Energy Sciences, Chemical Sciences, Geosciences, and Biosciences Division, through Contract No. DE-AC0276SF00515. 

\section*{Author contributions}
Conceptualization and methodology: X.L. with support from J.B.T., J.P.C. and P.J.H.; Dataset creation and curation: X.L. and P.J.H.; Generative model development: X.L., J.H., M.X. and S.E.; CEI experiments: X.L., T.J., R.B., M.M., M.N.P., D.R., A.R. and F.T.; CEI experiment data analysis: X.L. and T.J.; Original draft: X.L., T.J., R.B., J.H., M.X., D.R., F.T., T.J.A.W., S.E. and P.J.H.; Final draft: all authors.
\section*{Competing interests}
The authors declare no competing interests.
\section*{Data availability}
The datasets and model weights used in this study have been deposited in the Zenodo database under accession code 15794470 (https://zenodo.org/records/15794470) \cite{li_molexa_data_zenodo}.
\section*{Code availability}
The source code is publicly available at https://github.com/xli025/molexa \cite{li_molexa_code_zenodo}.

\clearpage
\section*{Supplementary Information}

\subsection*{Supplementary Figures}

\begin{suppfigure}[H] 
	\centering
	\includegraphics[width=1\textwidth]{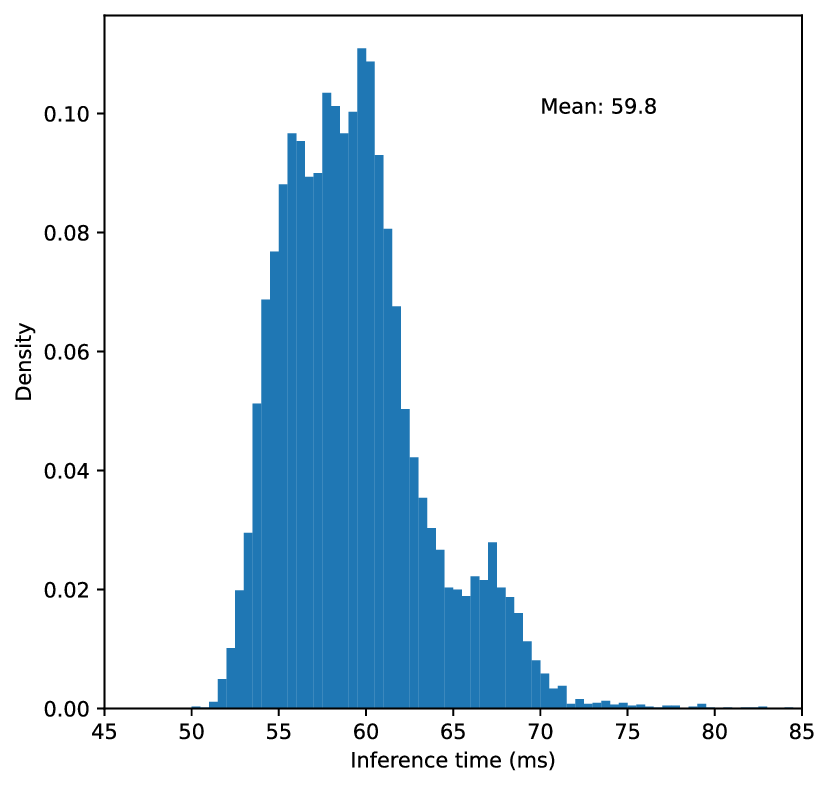} 

	\caption{\textbf{Inference time.}
		Distribution of the inference times for single-molecule reconstruction.}
	\label{fig:S_inference_time} 
\end{suppfigure}

\begin{suppfigure}[H] 
	\centering
	\includegraphics[width=1\textwidth]{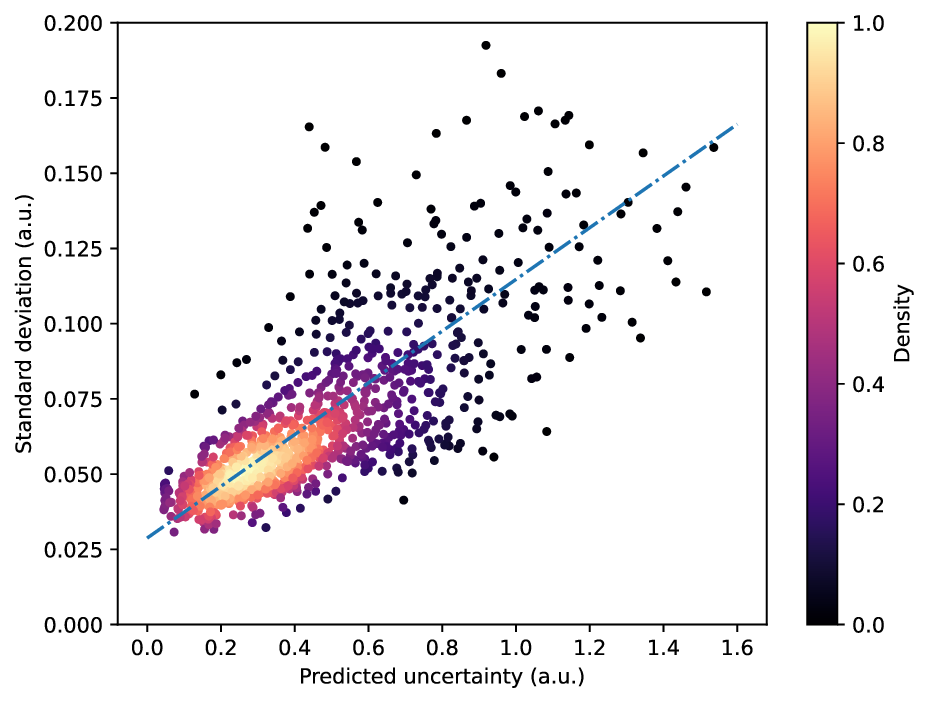} 

	\caption{\textbf{The relationship between the predicted uncertainty and the standard deviation of molecular-structure predictions.}
	The density plot was created with a thousand randomly selected input samples. For each sample, MOLEXA was run a thousand times, each producing a molecular structure with a predicted uncertainty attached to it. The standard deviation was calculated from these structure predictions. The plotted uncertainty was obtained by taking the average of the predicted uncertainties of each sample. The blue dot-dashed line shows the linear relationship (\(y = 0.086x\ +\ 0.029\)) between the predicted uncertainty and the standard deviation of the molecular-structure predictions.}
	\label{fig:S_uncertainty_std} 
\end{suppfigure}

\begin{suppfigure}[H] 
	\centering
	\includegraphics[width=1\textwidth]{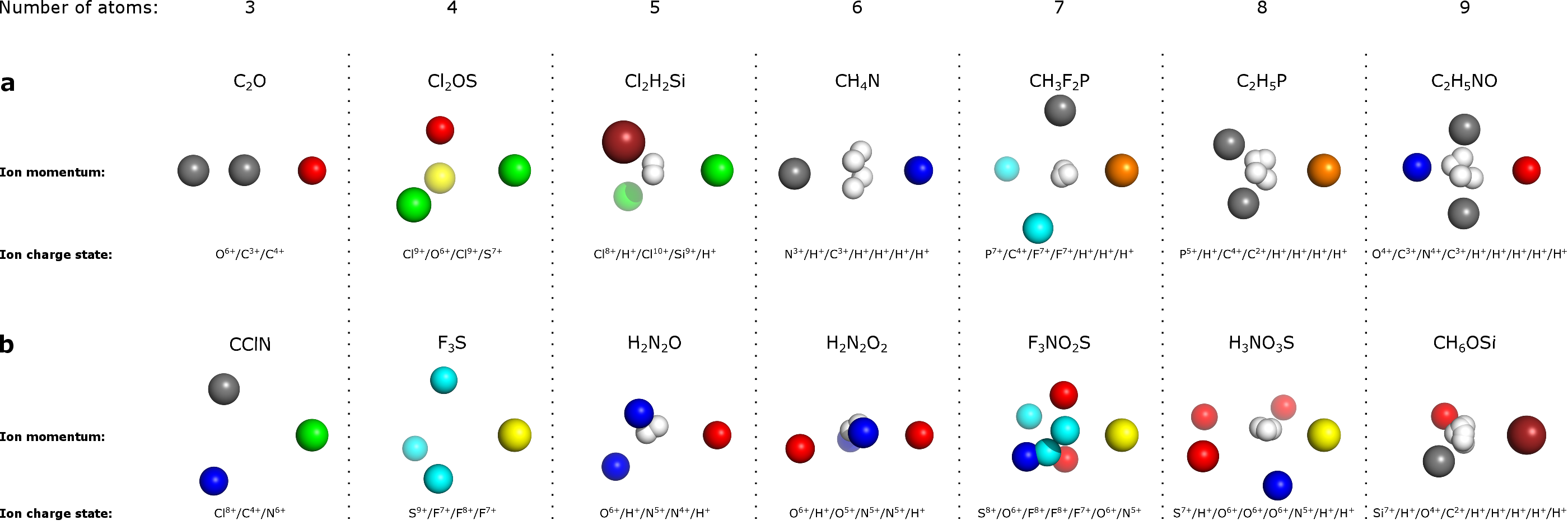} 

	\caption{\textbf{Input data for the structure predictions in Fig.~\ref{fig:performance}.}
		The ion-momentum distributions and charge states displayed in panels \textbf{a} and \textbf{b} are the model input for predicting the corresponding structures in Fig.~\ref{fig:performance}. The first two ions in the charge-state list are the ones used for defining the molecular frame.}
	\label{fig:S_momen_performance} 
\end{suppfigure}

\begin{suppfigure}[H] 
	\centering
	\includegraphics[width=1\textwidth]{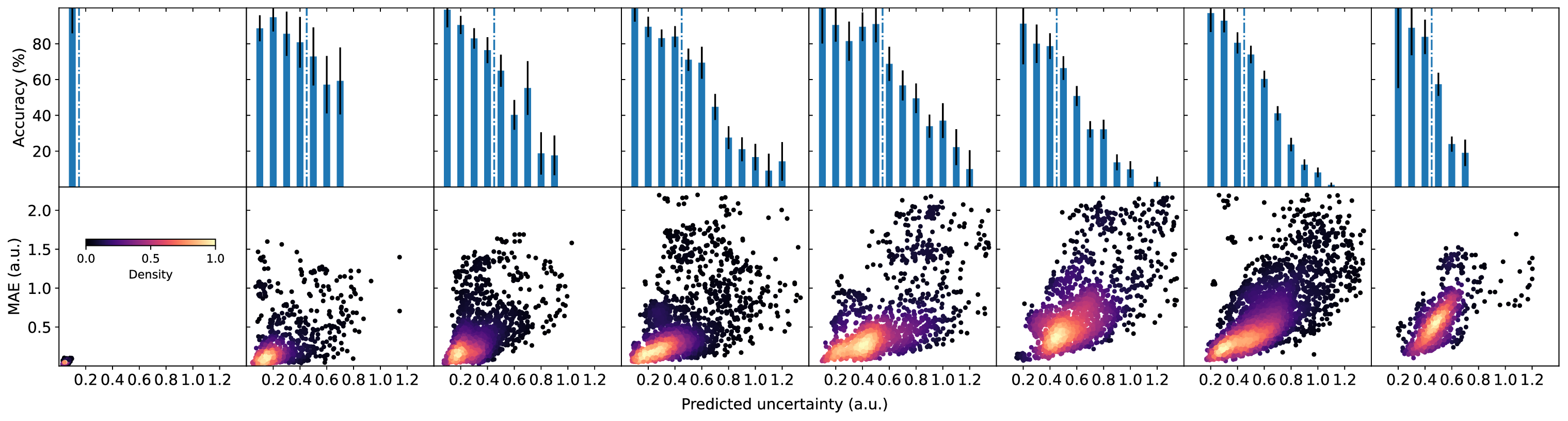} 
	\caption{\textbf{Overview of prediction accuracy, MAE and their dependence on predicted uncertainties.}
    Top row: The accuracy of the MOLEXA predictions as a function of uncertainty estimates. At each uncertainty bin of size 0.1 a.u., the bar height is the percentage of predictions with a MAE below 0.6~a.u out of all predictions which have an uncertainty belonging to that bin. The error bar is calculated from standard deviations, assuming a Poisson distribution for the number of counts falling into each bin. It can be seen from these plots that the accuracy generally drops with increasing uncertainty. For predictions with uncertainties smaller than the value marked by the dot-dashed vertical line in each subplot, it is expected that the accuracy of MOLEXA is greater than 75\%. Bottom row: The relation between the predicted uncertainty and the MAE of the reconstructed molecular structures. The density plots use the same color scale.
}
	\label{fig:S_accuracy_mae_uncertainty} 
\end{suppfigure}

\begin{suppfigure}[H] 
	\centering
	\includegraphics[width=1\textwidth]{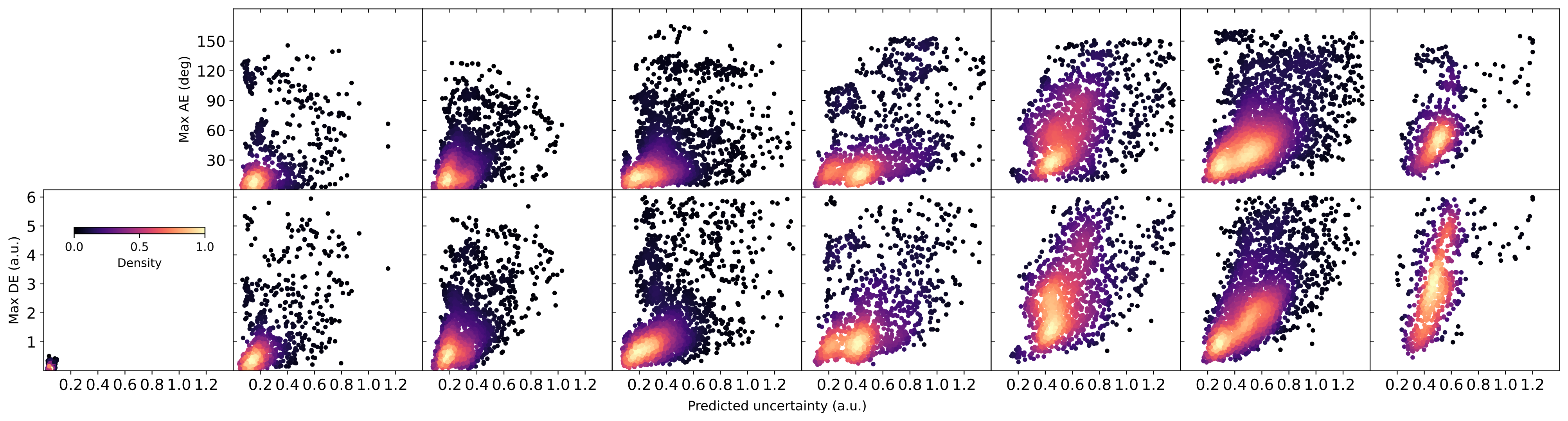} 

	\caption{\textbf{Overview of maximum atomic distance and angle errors.}
     Top row:  The dependence of the maximum angle error of the reconstructed molecular structures on the predicted uncertainty. Bottom row: The relation between the maximum distance error and the predicted uncertainty. The density plots use the same color scale.}
	\label{fig:S_max_DE_AE} 
\end{suppfigure}

\begin{suppfigure}[H] 
	\centering
	\includegraphics[width=0.95\textwidth]{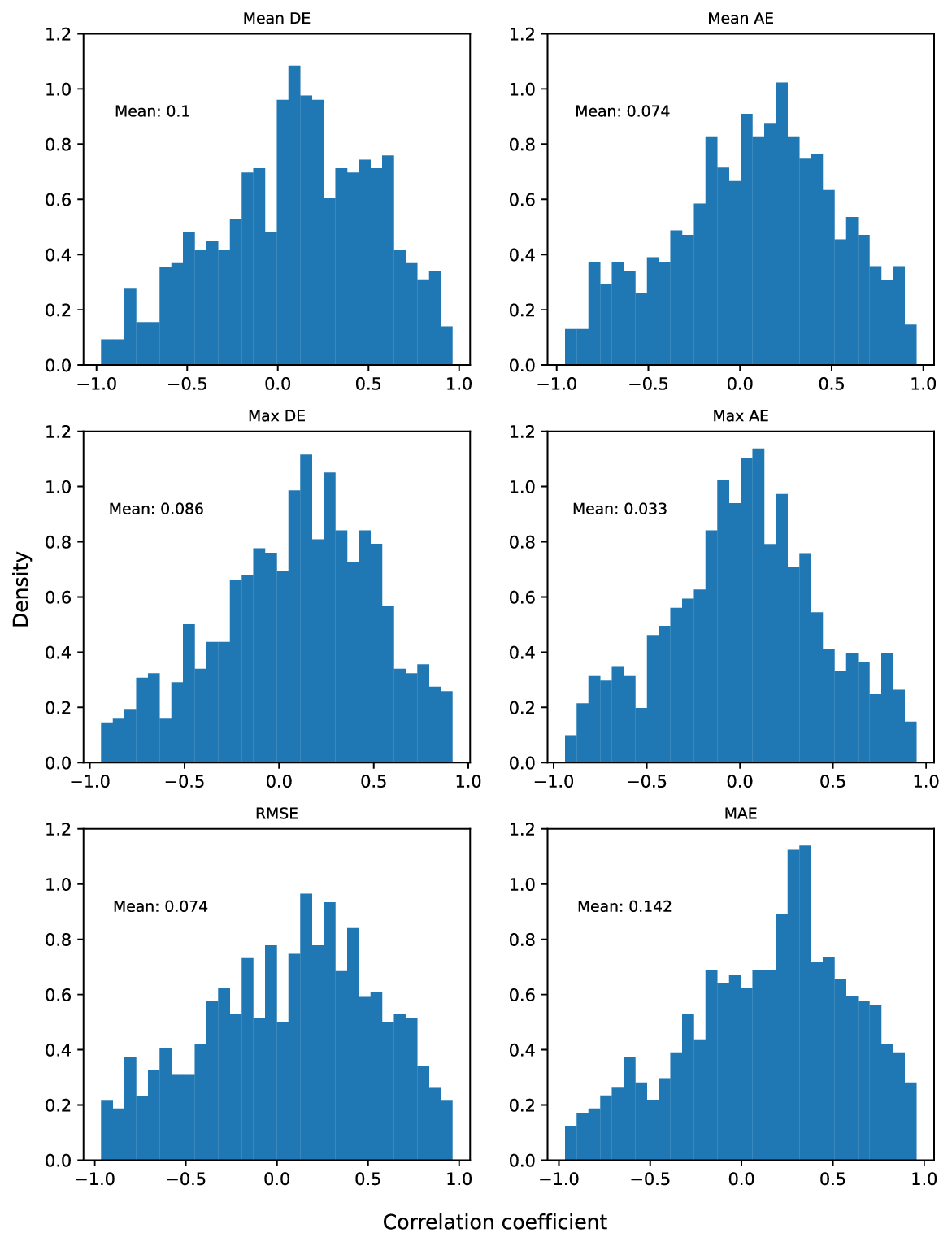} 

	\caption{\textbf{Correlation between reconstruction errors and predicted uncertainty.} Distributions of the correlation coefficients between reconstruction errors (top: mean distance and angle errors; middle: maximum distance and angle errors; bottom: root mean squared error and mean absolute error) and predicted uncertainties. The distributions were created with the same dataset as for Supplementary Fig.~\ref{fig:S_uncertainty_std}. The correlation coefficient was calculated as \(r = \frac{\sum(\mathbf{x}\ -\ m_x)(\mathbf{y}\ -\ m_y)}{\sqrt{\sum(\mathbf{x}\ -\ m_x)^2 \sum(\mathbf{y}\ -\ m_y)^2}}\), where \(m_x\) is the mean of \(\mathbf{x}\) array and \(m_y\) is the mean of \(\mathbf{y}\) array.}
	\label{fig:S_err_uncertainty_corr} 
\end{suppfigure}

\begin{suppfigure}[H] 
	\centering
	\includegraphics[width=0.9\textwidth]{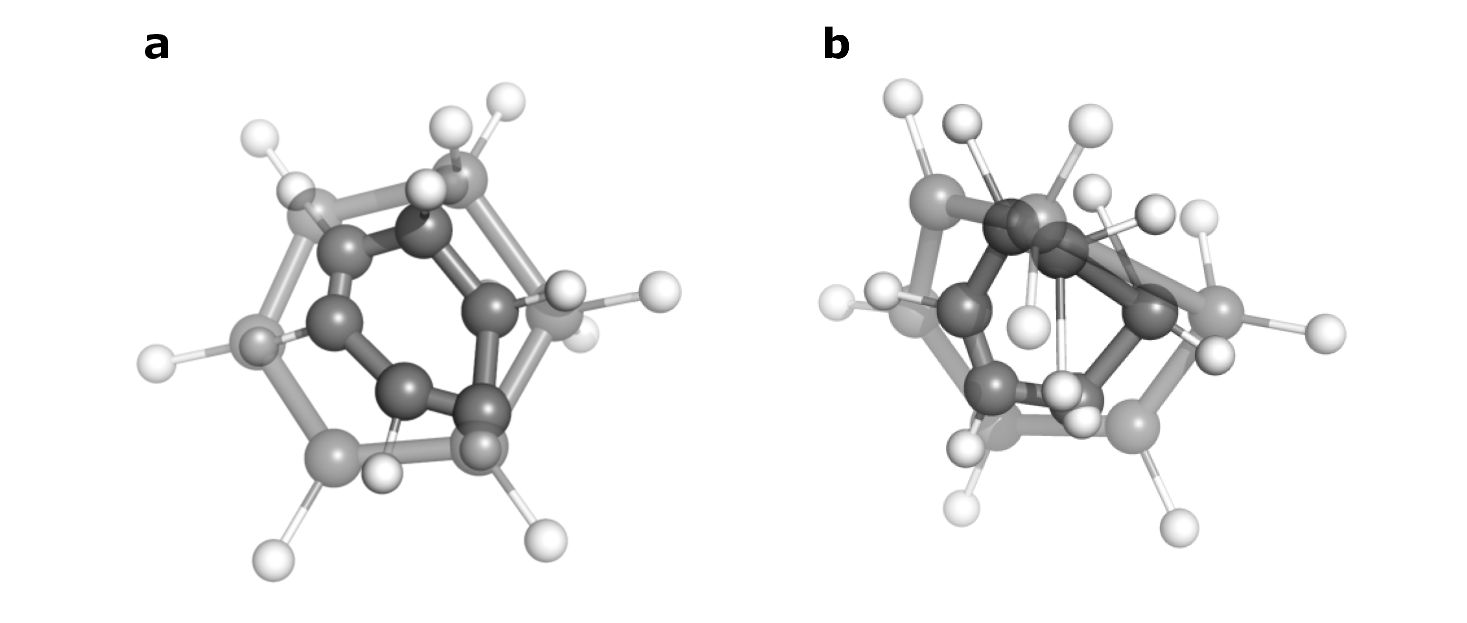} 

	\caption{\textbf{Prediction for larger molecules.}
		MOLEXA, which was trained with molecules containing up to seven atoms is expected to have difficulties predicting much larger molecules. The inaccurate MOLEXA predictions for the ground-state (\textbf{a}) and distorted (\textbf{b}) structures of the 1,3-cyclohexadiene molecule \cite{wolf_photochemical_2019} are displayed. The predicted and ground-truth structures are plotted as opaque and semi-transparent ball-and-stick models, respectively.}
	\label{fig:S_CHD} 
\end{suppfigure}

\begin{suppfigure}[H] 
	\centering
	\includegraphics[width=0.750\textwidth]{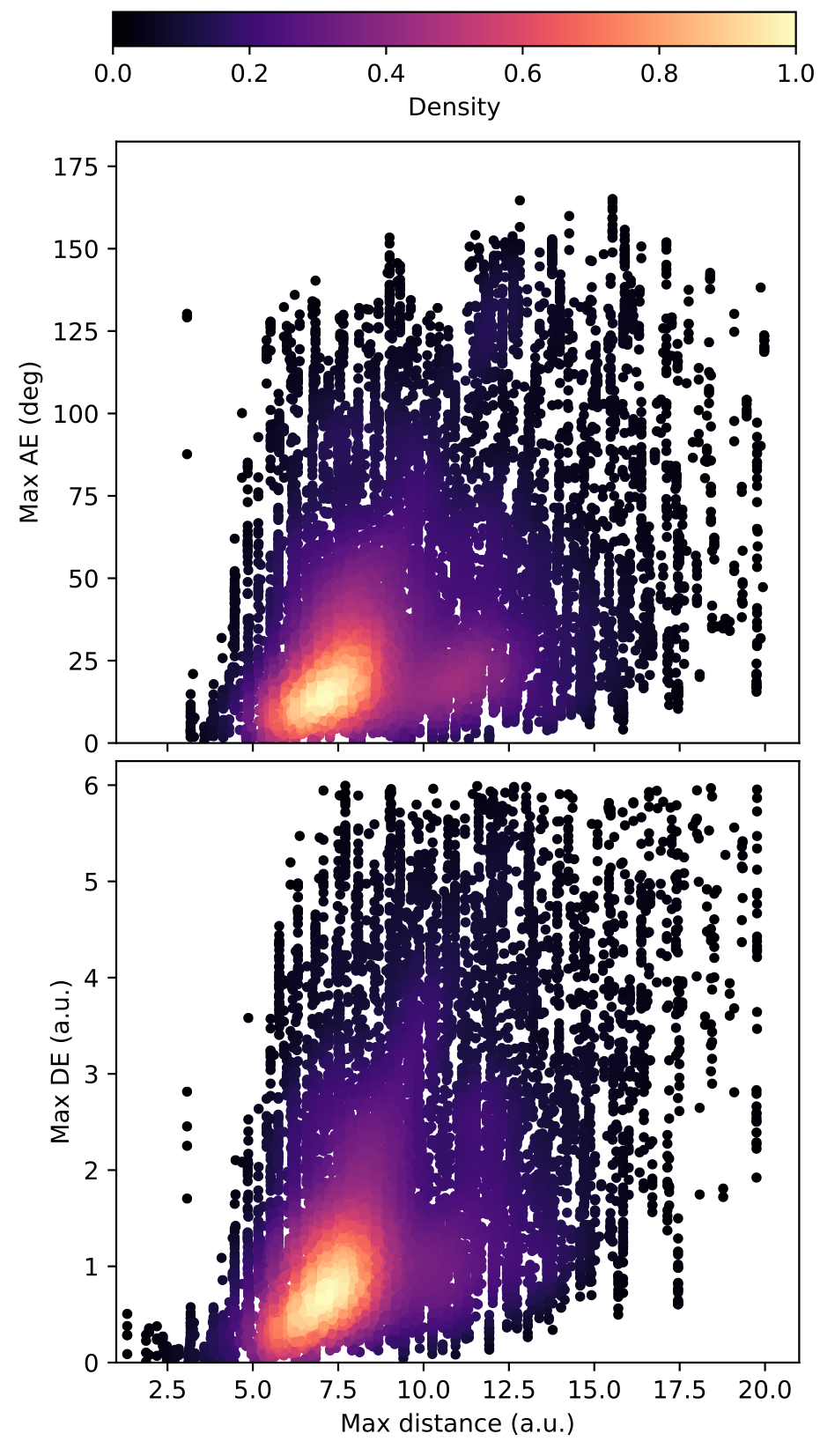} 

	\caption{\textbf{Dependence of reconstruction errors on atomic  distances.}
		The dependence of the maximum angle (top row) and distance (bottom row) errors on the maximum atomic distance in a molecule. The density plots use the same color scale.}
	\label{fig:S_DE_AE_distance} 
\end{suppfigure}

\begin{suppfigure}[H] 
	\centering
	\includegraphics[width=1\textwidth]{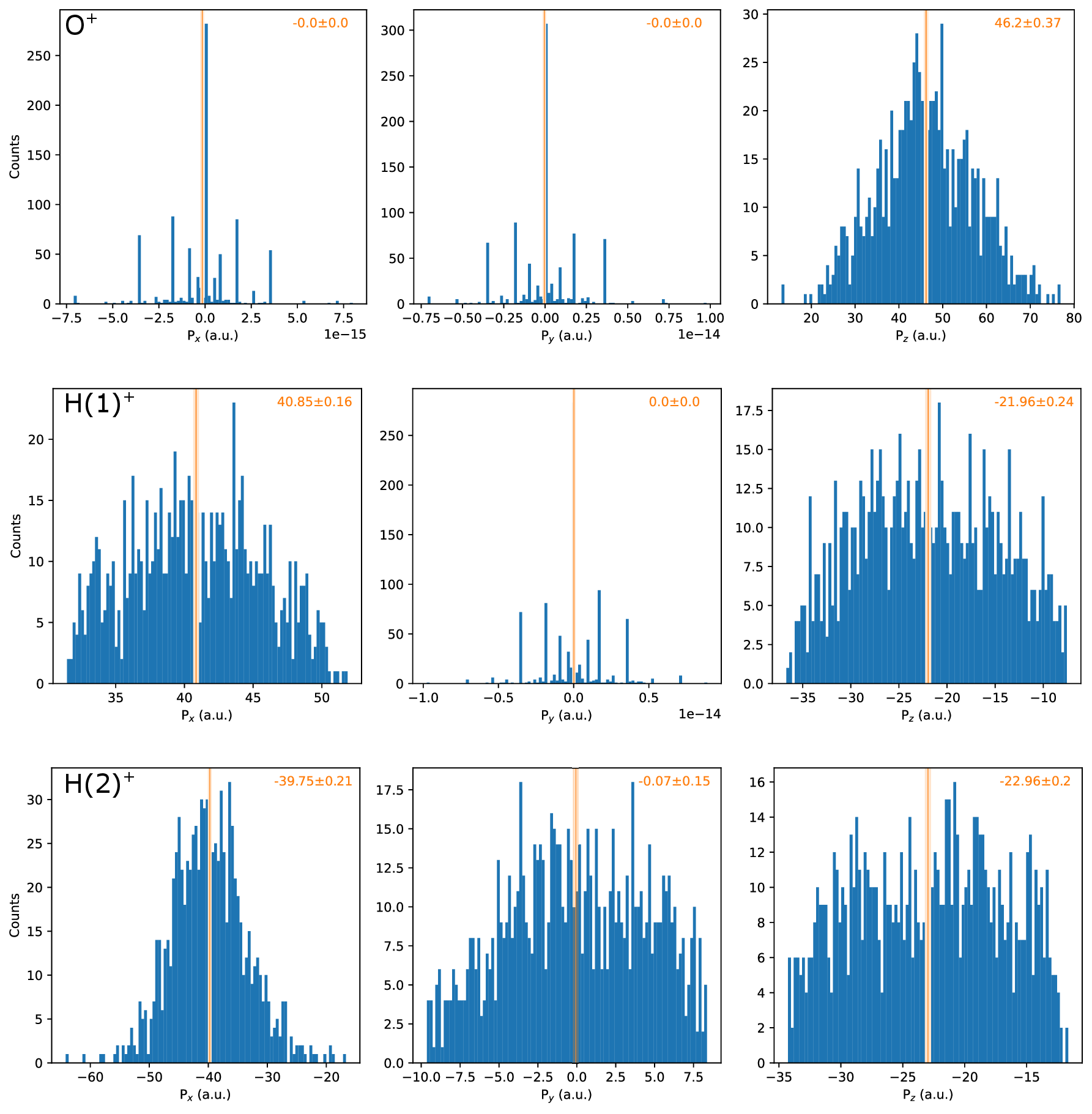} 

	\caption{\textbf{Experimental data of water.}
		The momentum components of O$^+$, H(1)$^+$, and H(2)$^+$ are shown as blue histograms. The centroid values are displayed in light orange.}
	\label{fig:S_H2O} 
\end{suppfigure}

\begin{suppfigure}[H] 
	\centering
	\includegraphics[width=1\textwidth]{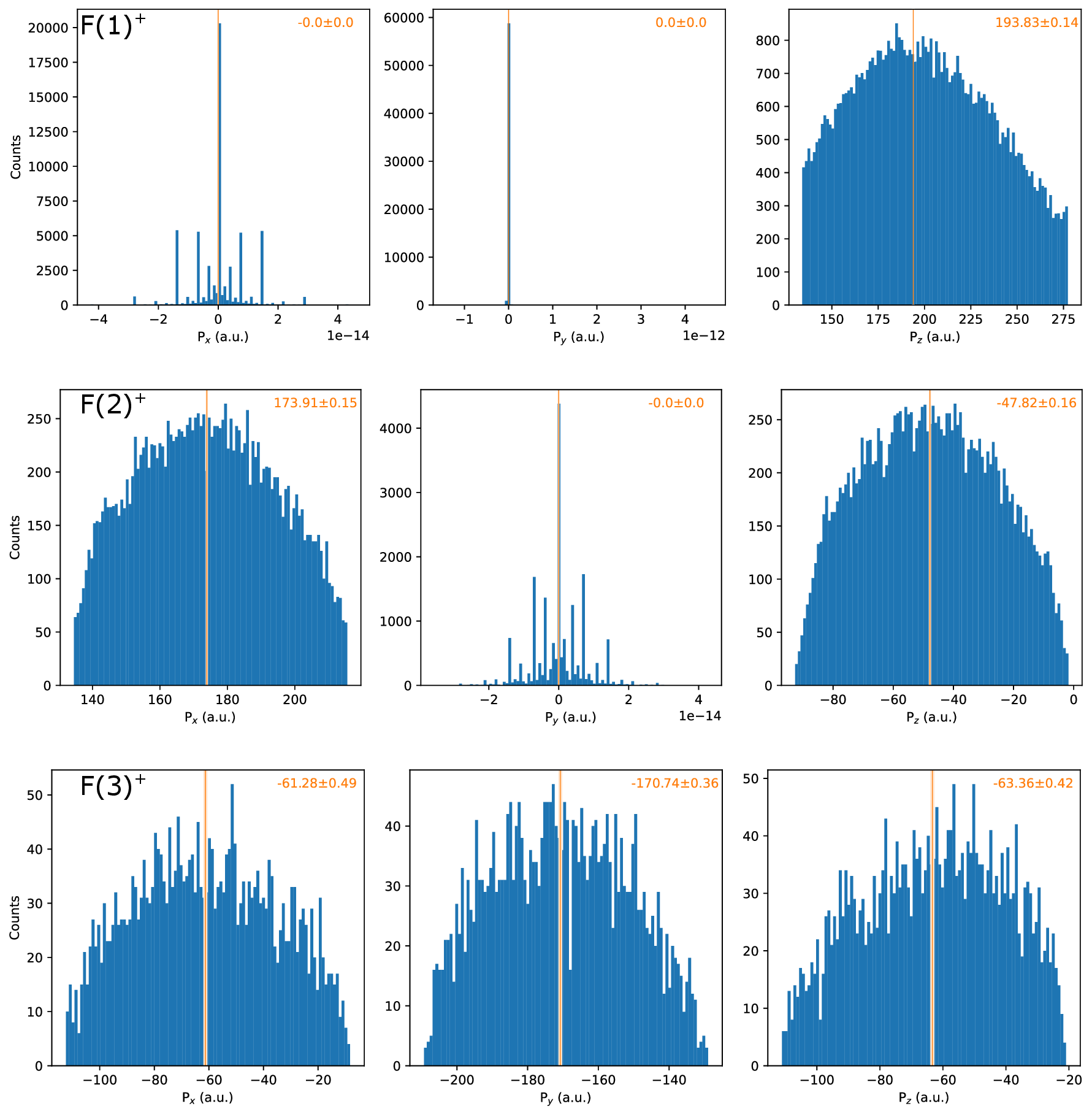} 

	\caption{\textbf{Experimental data of tetrafluoromethane (part 1).}
		The momentum components of F(1)$^+$, F(2)$^+$, and F(3)$^+$ are shown as blue histograms. The centroid values are displayed in light orange.}
	\label{fig:S_CF4_1} 
\end{suppfigure}

\begin{suppfigure}[H] 
	\centering
	\includegraphics[width=1\textwidth]{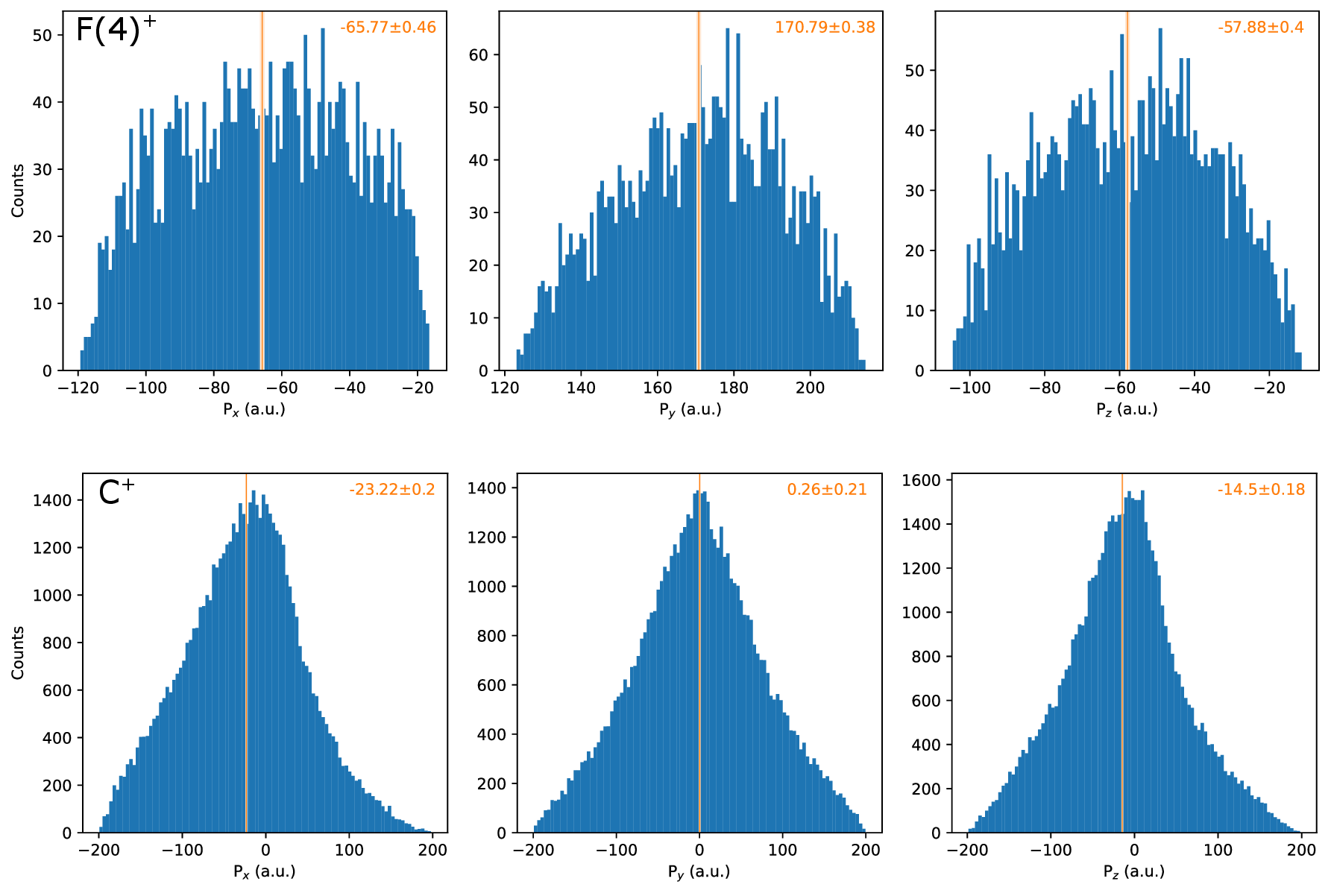} 

	\caption{\textbf{Experimental data of tetrafluoromethane (part 2).}
		The momentum components of F(4)$^+$ and C$^+$ are shown as blue histograms. The centroid values are displayed in light orange.}
	\label{fig:S_CF4_2} 
\end{suppfigure}

\begin{suppfigure}[H] 
	\centering
	\includegraphics[width=1\textwidth]{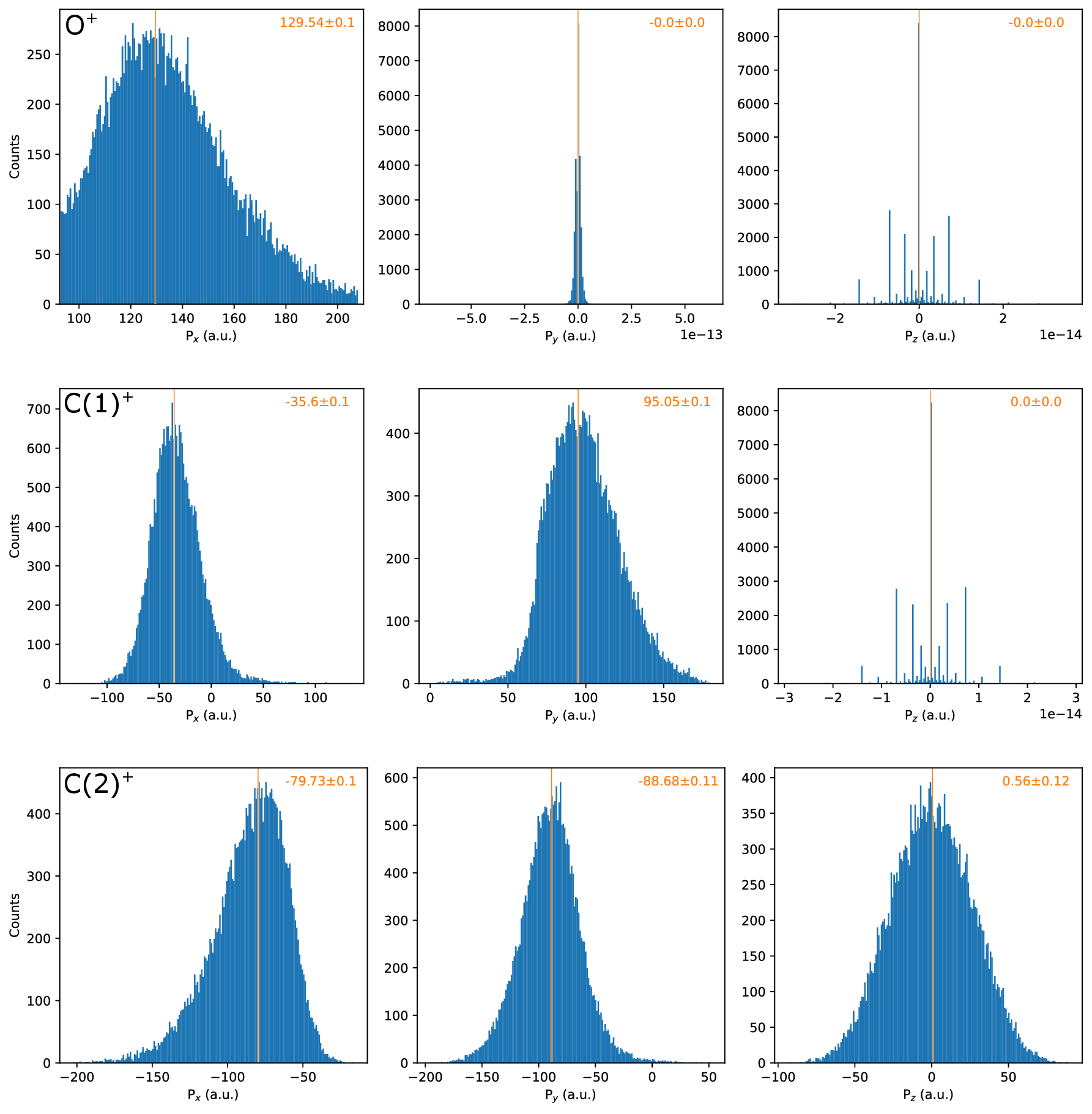} 

	\caption{\textbf{Experimental data of ethanol (part 1).}
		The momentum components of O$^+$, C(1)$^+$, and C(2)$^+$ are shown as blue histograms. The centroid values are displayed in light orange.}
	\label{fig:S_C2H6O_1} 
\end{suppfigure}

\begin{suppfigure}[H] 
	\centering
	\includegraphics[width=1\textwidth]{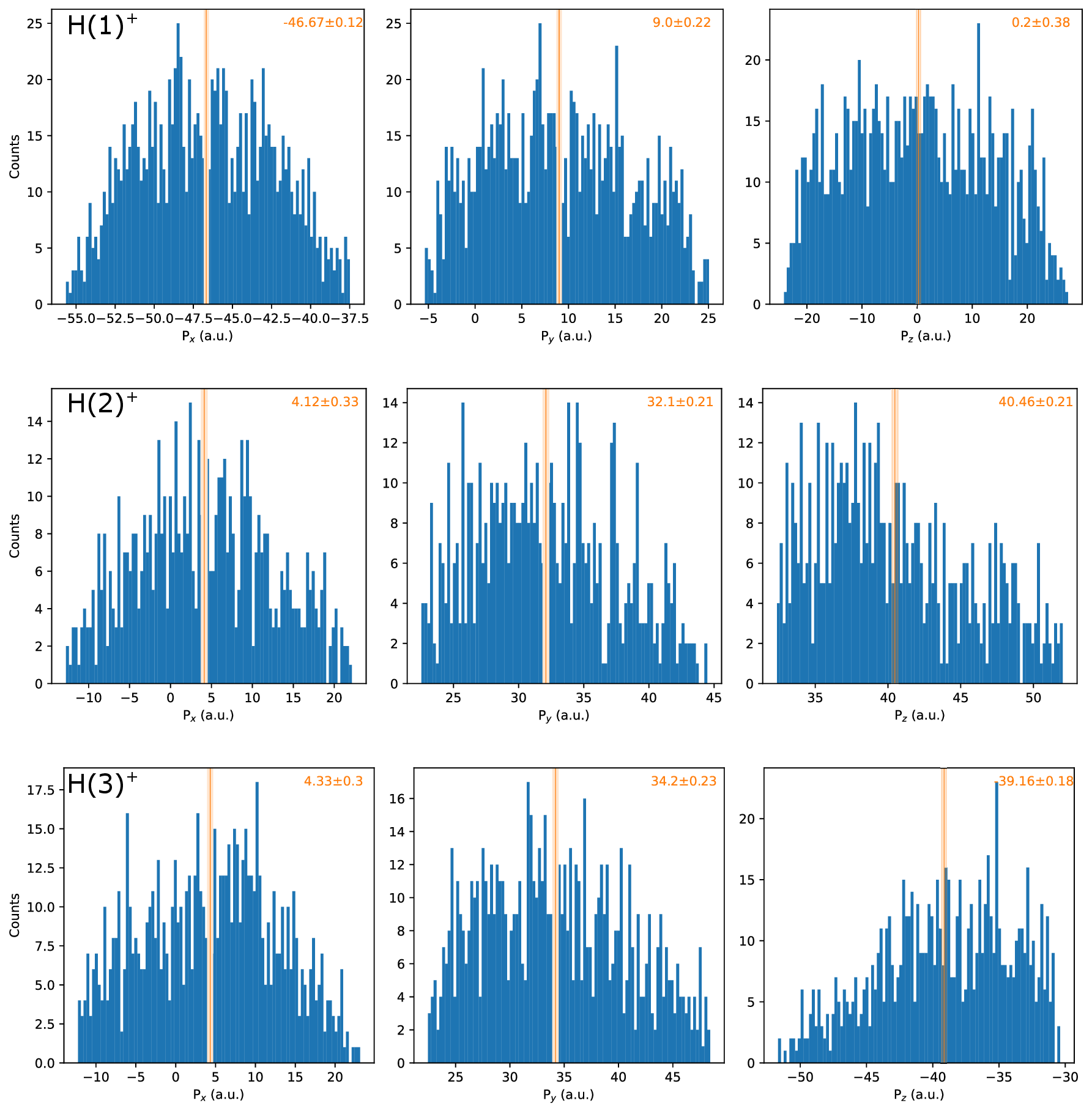} 

	\caption{\textbf{Experimental data of ethanol (part 2).}
		The momentum components of H(1)$^+$, H(2)$^+$, and H(3)$^+$ are shown as blue histograms. The centroid values are displayed in light orange.}
	\label{fig:S_C2H6O_2} 
\end{suppfigure}

\begin{suppfigure}[H] 
	\centering
	\includegraphics[width=1\textwidth]{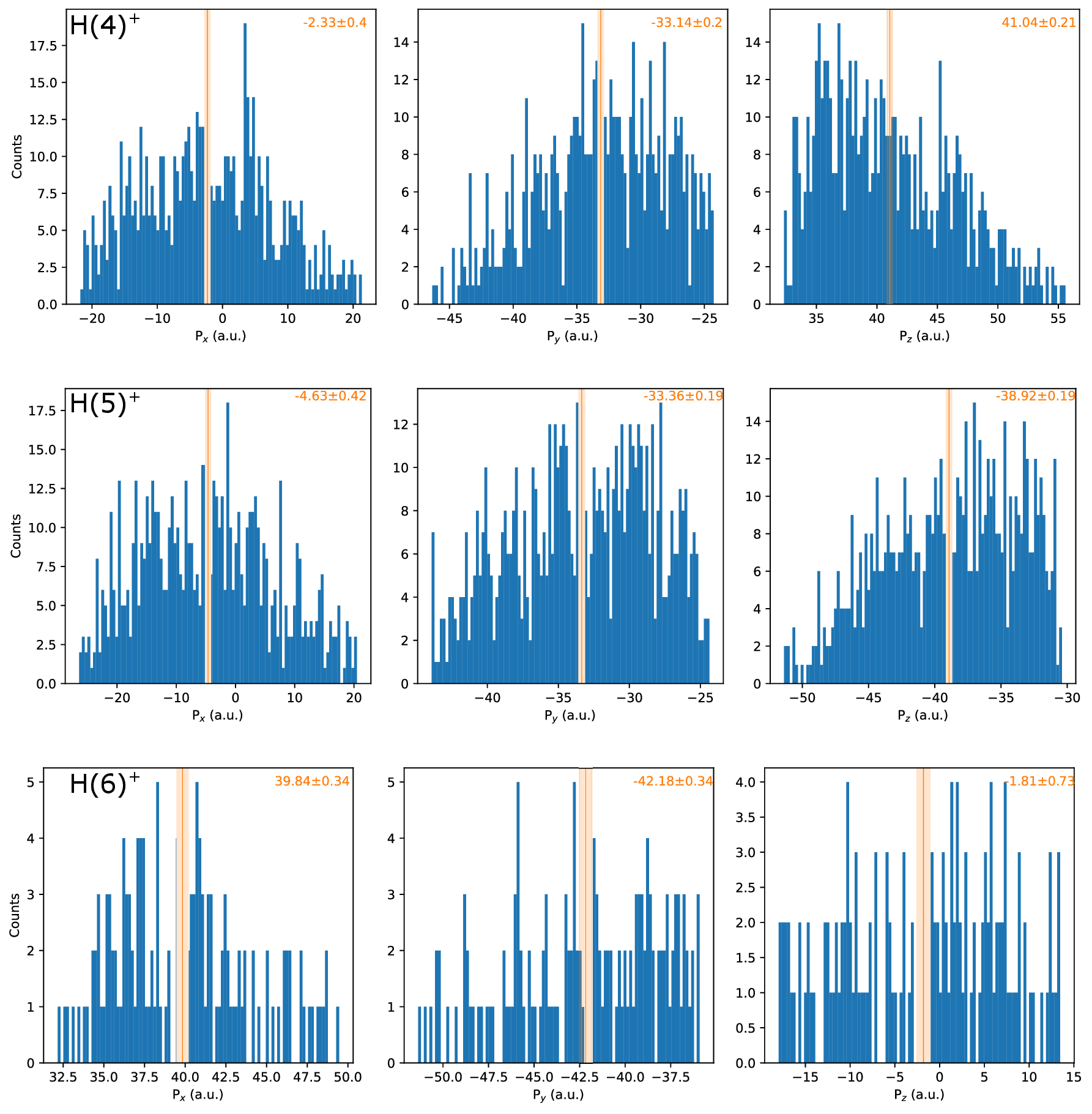} 

	\caption{\textbf{Experimental data of ethanol (part 3).}
		The momentum components of H(4)$^+$, H(5)$^+$, and H(6)$^+$ are shown as blue histograms. The centroid values are displayed in light orange.}
	\label{fig:S_C2H6O_3} 
\end{suppfigure}

\begin{suppfigure}[H] 
	\centering
	\includegraphics[width=0.9\textwidth]{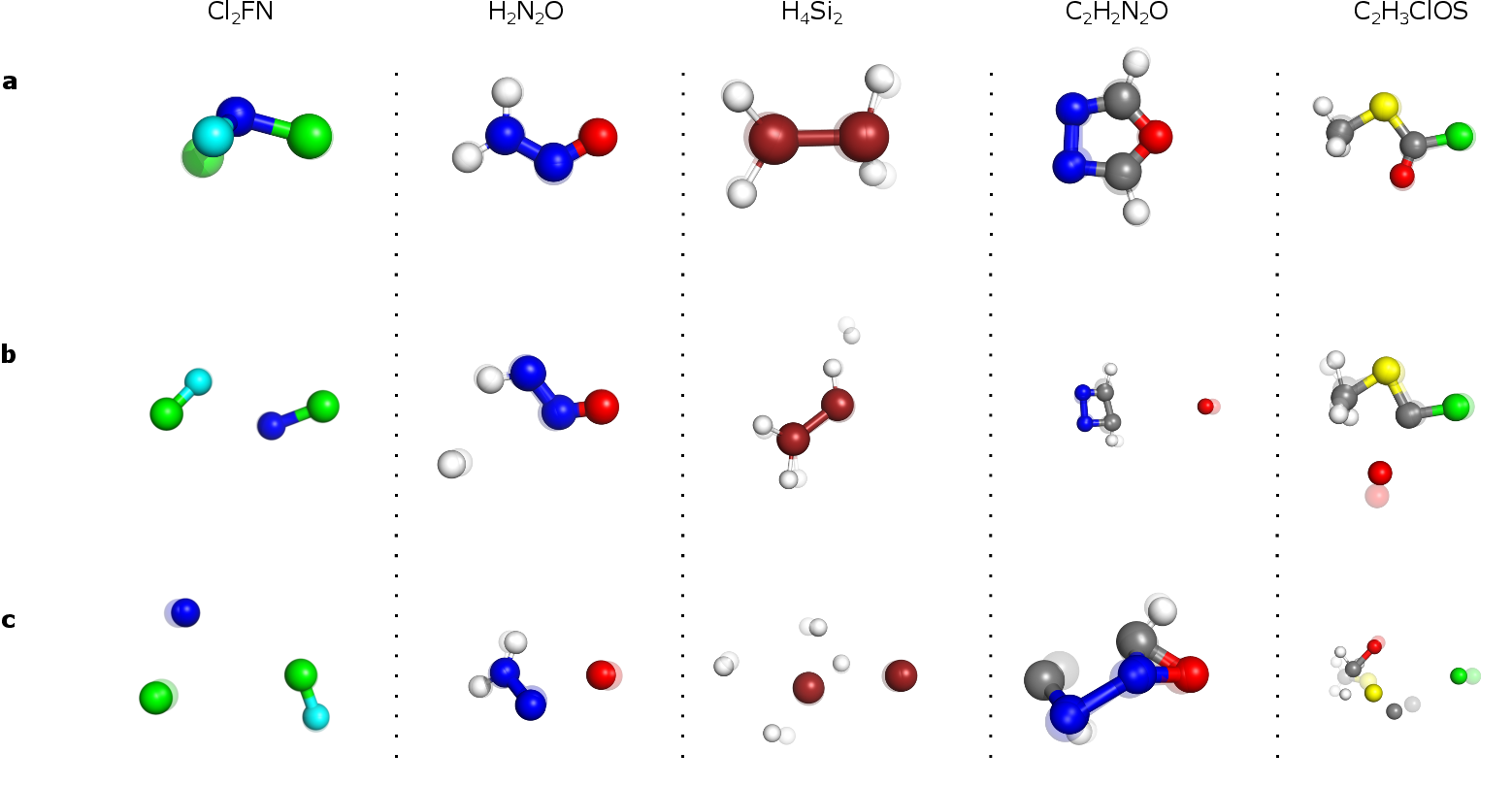} 

	\caption{\textbf{Exemplary MOLEXA predictions of changing molecular structures.} Predictions for three distinct structures of the same molecule are displayed in each column. \textbf{a,} Molecular structures in the ground state. \textbf{b} and \textbf{c,} Molecular structures differing from the ground state. The predicted and ground-truth structures are plotted as opaque and semi-transparent ball-and-stick models, respectively. The color coding of the elements is as follows - H: white, C: gray, N: blue, O: red, F: cyan, Si: brown, S: yellow, and Cl: green. The corresponding MOLEXA input data including the ion charge states and momentum distributions are shown in Supplementary Fig.~\ref{fig:S_momen_multi_structures}.}
	\label{fig:S_multi_structures} 
\end{suppfigure}

\begin{suppfigure}[H] 
	\centering
	\includegraphics[width=1\textwidth]{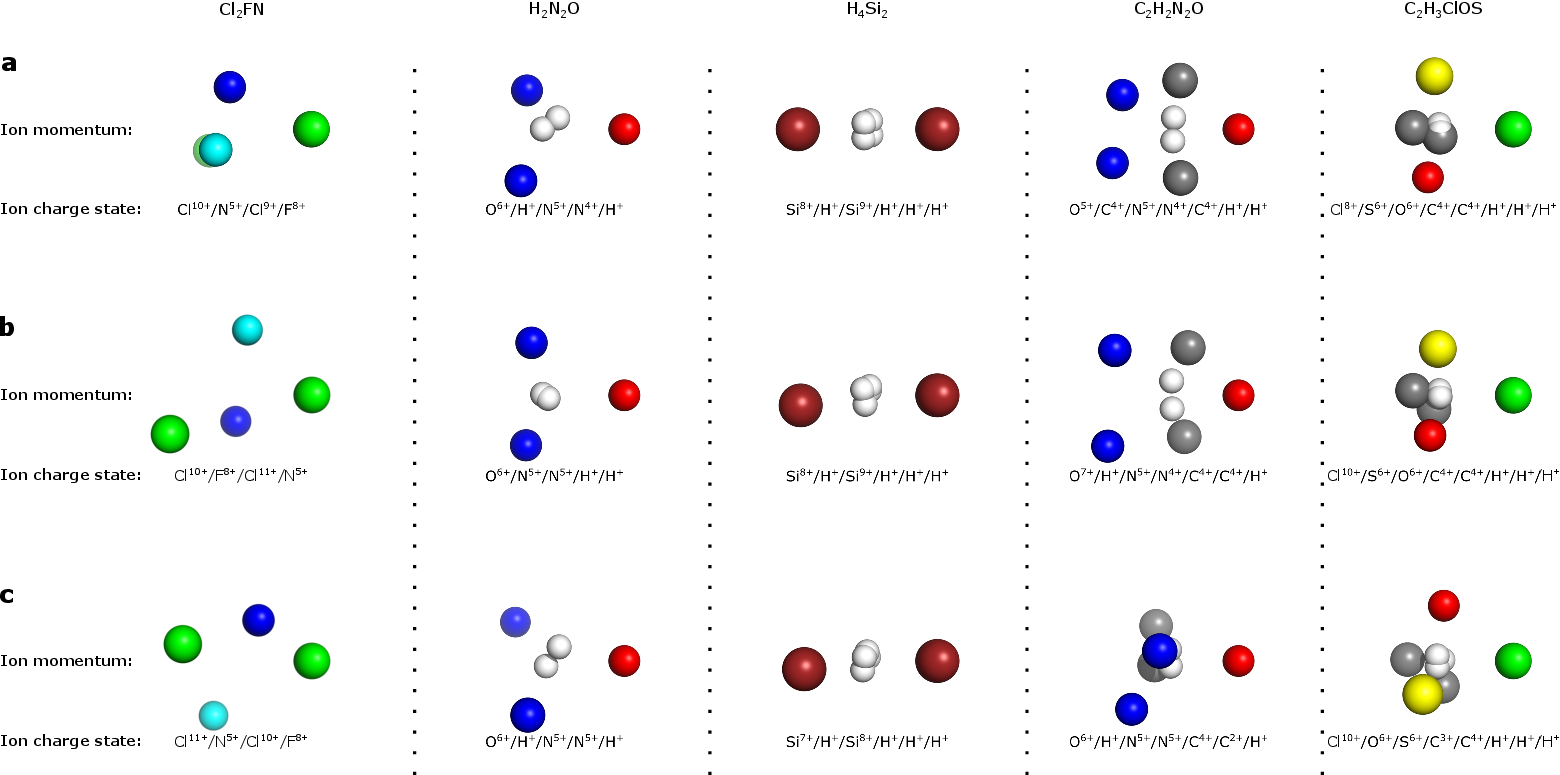} 

	\caption{\textbf{Input data for the structure predictions in Fig.~\ref{fig:S_multi_structures}.}
		The ion-momentum distribution and charge states displayed in panels \textbf{a}, \textbf{b}, and \textbf{c} are the model input for predicting the corresponding structures in Supplementary Fig.~\ref{fig:S_multi_structures}. The first two ions in the charge-state list are the ones used for defining the molecular frame.}
	\label{fig:S_momen_multi_structures} 
\end{suppfigure}

\begin{suppfigure}[H] 
	\centering
	\includegraphics[width=1\textwidth]{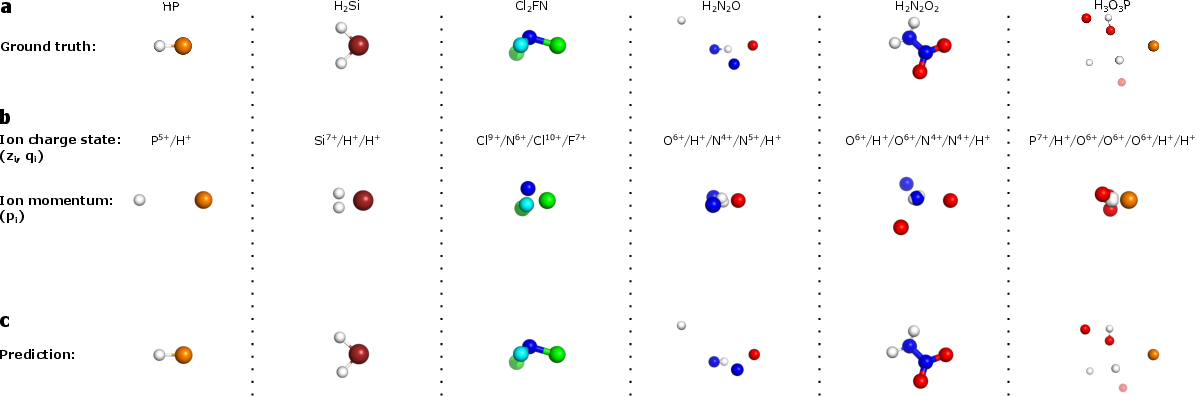} 

	\caption{\textbf{Data samples and predictions.}
		\textbf{a,} The ground-truth structure of different molecules with increasing sizes from left to right. \textbf{b,} The input data to MOLEXA, which includes the atomic number, charge state, and momentum distribution of the ion fragments from Coulomb explosion. The first two ions in the charge-state list are the ones used for defining the molecular frame. \textbf{c,} The retrieved molecular structures.}
	\label{fig:S_samples_predictions} 
\end{suppfigure}

\begin{suppfigure}[H] 
	\centering
	\includegraphics[width=1\textwidth]{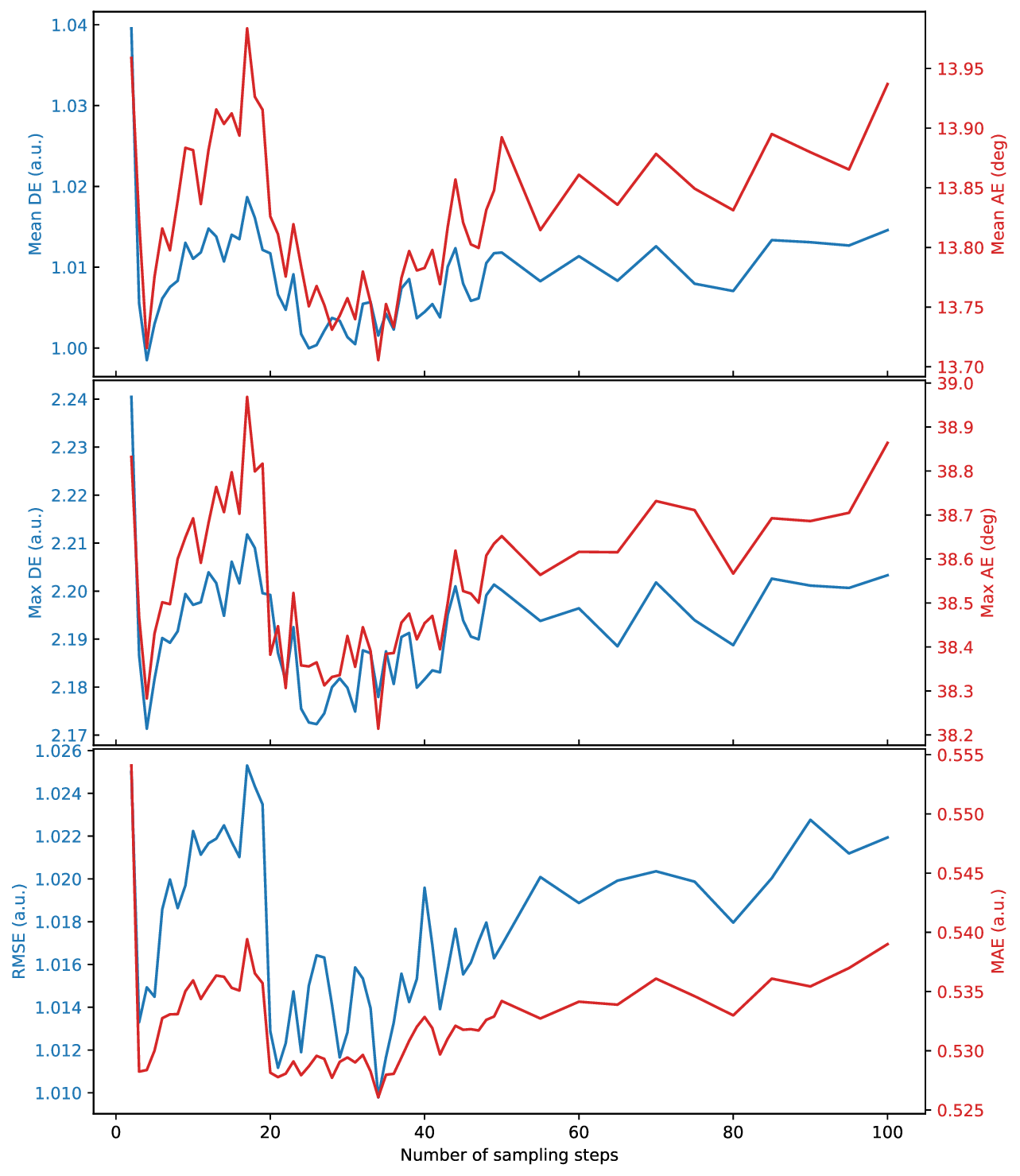} 

	\caption{\textbf{Reconstruction errors as a function of the number of sampling steps.} The dependence of reconstruction errors (top: mean distance and angle errors; middle: maximum distance and angle errors; bottom: root mean squared error and mean absolute error) on the number of sampling steps.}
	\label{fig:S_err_nfe} 
\end{suppfigure}

\clearpage
\subsection*{Supplementary Tables}

\begin{supptable}[h] 
	\centering
	\caption{\textbf{Two-stage training}}
	\label{tab:training} 
    \scalebox{0.8}{
	\begin{tabular}{|c|c|c|c|c|} 
		\hline
		\textbf{Training stage} & \textbf{Training samples} & \textbf{Batch size} & \textbf{Training time (hours)} & \textbf{GPUs (A100)}\\
		\hline
		1 & 5.7 $\times$ 10$^{6}$ & 32 & 82 & 16\\
        \hline
		2 & 6.1 $\times$ 10$^{4}$ & 16 & 1 & 4\\
		\hline
	\end{tabular}
    }
\end{supptable}

\begin{supptable}[h]
    \centering
    \caption{\textbf{Maximum distance and angle error values for the low-uncertainty predictions in Fig.~\ref{fig:performance}}}
    \label{tab:max_de_ae_low}

    \begin{tabular}{|l|c|c|}
        \hline
        \textbf{Molecule} & \textbf{Max DE (a.u.)} & \textbf{Max AE (deg)} \\
        \hline
        C$_2$O      & 0.083 & 4.293 \\
        \hline
        Cl$_2$OS    & 0.341 & 11.815 \\
        \hline
        Cl$_2$H$_2$Si & 0.341 & 3.072 \\
        \hline
        CH$_4$N     & 0.579 & 8.760 \\
        \hline
        CH$_3$F$_2$P & 0.552 & 6.392 \\
        \hline
        C$_2$H$_5$P & 0.422 & 16.640 \\
        \hline
        C$_2$H$_5$NO & 1.007 & 27.774 \\
        \hline
    \end{tabular}
\end{supptable}

\begin{supptable}[h]
    \centering
    \caption{\textbf{Maximum distance and angle error values for the high-uncertainty predictions in Fig.~\ref{fig:performance}}}
    \label{tab:max_de_ae_high}

    \begin{tabular}{|l|c|c|}
        \hline
        \textbf{Molecule} & \textbf{Max DE (a.u.)} & \textbf{Max AE (deg)} \\
        \hline
        CClN       & 3.044 & 60.729 \\
        \hline
        F$_3$S     & 3.121 & 51.317 \\
        \hline
        H$_2$N$_2$O & 3.807 & 68.760 \\
        \hline
        H$_2$N$_2$O$_2$ & 3.771 & 44.218 \\
        \hline
        F$_3$NO$_2$S & 4.372 & 69.623 \\
        \hline
        H$_3$NO$_3$S & 3.139 & 64.383 \\
        \hline
        CH$_6$OSi  & 3.924 & 76.091 \\
        \hline
    \end{tabular}
\end{supptable}

\begin{supptable}[h]
  \centering
  \caption{\textbf{Beamtime DOI and X‑ray pulse parameters}}
  \label{tab:exp}

  \renewcommand{\arraystretch}{1.1}
  \setlength{\tabcolsep}{6pt}

  \scalebox{0.75}{%
    \begin{tabular}{|c|c|c|c|}
      \hline
      \textbf{Molecule} & \textbf{DOI} & \textbf{Photon energy (eV)} & \textbf{Pulse energy (mJ)} \\
      \hline
      Water              & \href{https://doi.org/10.22003/XFEL.EU-DATA-002150-00}{10.22003/XFEL.EU-DATA-002150-00} & 1000 & 4.0  \\
      \hline
      Tetrafluoromethane & \href{https://doi.org/10.22003/XFEL.EU-DATA-002181-00}{10.22003/XFEL.EU-DATA-002181-00} & 1200 & 1.00 \\
      \hline
      Ethanol            & \href{https://doi.org/10.22003/XFEL.EU-DATA-002926-00}{10.22003/XFEL.EU-DATA-002926-00} & 1200 & 1.3  \\
      \hline
    \end{tabular}%
  }
\end{supptable}

\begin{supptable}[h]
    \centering
    \caption{\textbf{Water}}
    \label{tab:h2o}

    \begin{tabular}{|l|c|c|c|}
        \hline
        \textbf{Atom} & \textbf{X (a.u.)} & \textbf{Y (a.u.)} & \textbf{Z (a.u.)} \\ 
        \hline
        \multicolumn{4}{|l|}{\textbf{Prediction:}} \\ 
        \hline
        O & $0.594 \pm 0.248$ & $0.083 \pm 0.000$ & $0.049 \pm 0.000$ \\
        \hline
        H & $-0.432 \pm 0.243$ & $1.985 \pm 0.196$ & $-0.039 \pm 0.054$ \\
        \hline
        H & $-0.162 \pm 0.251$ & $-2.067 \pm 0.125$ & $-0.011 \pm 0.047$ \\
        \hline
        \multicolumn{4}{|l|}{\textbf{Ground truth:}} \\ 
        \hline
        O & $0.739$ & $0.000$ & $0.000$ \\
        \hline
        H & $-0.370$ & $1.431$ & $0.000$ \\
        \hline
        H & $-0.370$ & $-1.431$ & $0.000$ \\
        \hline
    \end{tabular}
\end{supptable}

\newpage

\begin{supptable}[h]
    \centering
    \caption{\textbf{Tetrafluoromethane}}
    \label{tab:cf4}

    \begin{tabular}{|l|c|c|c|}
        \hline
        \textbf{Atom} & \textbf{X (a.u.)} & \textbf{Y (a.u.)} & \textbf{Z (a.u.)} \\ 
        \hline
        \multicolumn{4}{|l|}{\textbf{Prediction:}} \\ 
        \hline
        F & $2.223 \pm 0.516$  & $0.026 \pm 0.047$  & $-0.001 \pm 0.050$ \\ 
        \hline
        F & $-0.486 \pm 0.832$ & $1.821 \pm 0.441$  & $0.012 \pm 0.000$  \\ 
        \hline
        F & $-0.610 \pm 0.769$ & $-0.698 \pm 0.514$ & $-1.505 \pm 0.384$ \\ 
        \hline
        F & $-0.890 \pm 0.791$ & $-0.982 \pm 0.467$ & $1.665 \pm 0.469$  \\ 
        \hline
        C & $-0.237 \pm 0.645$ & $-0.166 \pm 0.361$ & $-0.171 \pm 0.488$ \\ 
        \hline
        \multicolumn{4}{|l|}{\textbf{Ground truth:}} \\ 
        \hline
        F & $2.486$      & $-0.000$      & $0.000$      \\ 
        \hline
        F & $-0.829$     & $2.344$       & $0.000$      \\ 
        \hline
        F & $-0.829$     & $-1.172$      & $-2.030$     \\ 
        \hline
        F & $-0.829$     & $-1.172$      & $2.030$      \\ 
        \hline
        C & $0.000$      & $0.000$       & $0.000$      \\ 
        \hline
    \end{tabular}
\end{supptable}

\newpage

\begin{supptable}[h]
    \centering
    \caption{\textbf{Ethanol}}
    \label{tab:c2h6o}

    \begin{tabular}{|l|c|c|c|}
        \hline
        \textbf{Atom} & \textbf{X (a.u.)} & \textbf{Y (a.u.)} & \textbf{Z (a.u.)} \\ 
        \hline
        \multicolumn{4}{|l|}{\textbf{Prediction:}} \\ 
        \hline
        O & $2.556 \pm 1.505$  & $0.229 \pm 0.252$   & $-0.046 \pm 0.000$  \\
        \hline
        H & $0.730 \pm 1.218$  & $3.450 \pm 0.621$   & $-0.044 \pm 0.000$  \\
        \hline
        C & $-1.105 \pm 0.941$ & $-1.182 \pm 0.832$  & $-1.046 \pm 0.638$  \\
        \hline
        C & $-0.161 \pm 0.956$ & $1.294 \pm 0.607$   & $0.859 \pm 0.469$   \\
        \hline
        H & $4.848 \pm 1.307$  & $1.076 \pm 0.566$   & $0.810 \pm 0.585$   \\
        \hline
        H & $-4.199 \pm 1.588$ & $-0.332 \pm 0.843$  & $0.263 \pm 0.876$   \\
        \hline
        H & $-1.484 \pm 1.563$ & $-1.606 \pm 0.831$  & $-3.645 \pm 0.682$  \\
        \hline
        H & $-1.280 \pm 1.346$ & $-4.021 \pm 0.666$  & $-0.524 \pm 0.571$  \\
        \hline
        H & $0.096 \pm 1.940$  & $1.091 \pm 0.870$   & $3.372 \pm 0.671$   \\
        \hline
        \multicolumn{4}{|l|}{\textbf{Ground truth:}} \\ 
        \hline
        O & $2.816$      & $-0.000$     & $0.000$      \\
        \hline
        H & $0.051$      & $2.698$      & $0.000$      \\
        \hline
        C & $-1.490$     & $-1.011$     & $-0.799$     \\
        \hline
        C & $0.285$      & $0.746$      & $0.590$      \\
        \hline
        H & $3.894$      & $1.165$      & $0.921$      \\
        \hline
        H & $-3.432$     & $-0.439$     & $-0.347$     \\
        \hline
        H & $-1.088$     & $-0.851$     & $-2.809$     \\
        \hline
        H & $-1.088$     & $-2.930$     & $-0.180$     \\
        \hline
        H & $0.051$      & $0.623$      & $2.625$      \\
        \hline
    \end{tabular}
\end{supptable}

\begin{supptable}[h]
    \centering
    \caption{\textbf{Cyclobutane, S$_0$}}
    \label{tab:c4h6_s0}

    \begin{tabular}{|l|c|c|c|}
        \hline
        \textbf{Atom} & \textbf{X (a.u.)} & \textbf{Y (a.u.)} & \textbf{Z (a.u.)} \\ 
        \hline
        \multicolumn{4}{|l|}{\textbf{Prediction:}} \\ 
        \hline
        C & $2.751 \pm 0.452$   & $0.016 \pm 0.000$   & $-0.008 \pm 0.000$  \\ 
        \hline
        C & $0.137 \pm 0.410$   & $1.398 \pm 0.591$   & $-0.011 \pm 0.000$  \\ 
        \hline
        C & $0.237 \pm 0.381$   & $-1.429 \pm 0.507$  & $-0.180 \pm 0.255$  \\ 
        \hline
        C & $-2.372 \pm 0.584$  & $-0.485 \pm 0.358$  & $-0.218 \pm 0.194$  \\ 
        \hline
        H & $5.194 \pm 0.888$   & $-0.002 \pm 0.276$  & $0.407 \pm 0.168$   \\ 
        \hline
        H & $0.187 \pm 0.945$   & $-3.205 \pm 0.415$  & $0.152 \pm 0.273$   \\ 
        \hline
        H & $0.314 \pm 0.712$   & $2.745 \pm 0.471$   & $-1.451 \pm 0.310$  \\ 
        \hline
        H & $-3.408 \pm 0.549$  & $-0.986 \pm 0.394$  & $1.316 \pm 0.226$   \\ 
        \hline
        H & $0.006 \pm 0.652$   & $2.170 \pm 0.549$   & $2.008 \pm 0.284$   \\ 
        \hline
        H & $-3.046 \pm 0.541$  & $-0.222 \pm 0.467$  & $-2.014 \pm 0.270$  \\ 
        \hline
        \multicolumn{4}{|l|}{\textbf{Ground truth:}} \\ 
        \hline
        C & $2.008$      & $0.000$       & $0.000$      \\ 
        \hline
        C & $-0.062$     & $1.857$       & $0.000$      \\ 
        \hline
        C & $0.689$      & $-2.015$      & $-0.187$     \\ 
        \hline
        C & $-1.709$     & $-0.539$      & $-0.181$     \\ 
        \hline
        H & $4.219$      & $0.210$       & $0.375$      \\ 
        \hline
        H & $0.738$      & $-4.185$      & $0.225$      \\ 
        \hline
        H & $0.049$      & $3.150$       & $-1.448$     \\ 
        \hline
        H & $-3.093$     & $-0.942$      & $1.292$      \\ 
        \hline
        H & $-0.425$     & $2.895$       & $2.084$      \\ 
        \hline
        H & $-2.413$     & $-0.431$      & $-2.161$     \\ 
        \hline
    \end{tabular}
\end{supptable}

\begin{supptable}[h]
    \centering
    \caption{\textbf{Cyclobutane, S$_1$/S$_0$ Minimum Energy Conical Intersection (MECI)}}
    \label{tab:c4h6_meci}

    \begin{tabular}{|l|c|c|c|}
        \hline
        \textbf{Atom} & \textbf{X (a.u.)} & \textbf{Y (a.u.)} & \textbf{Z (a.u.)} \\ 
        \hline
        \multicolumn{4}{|l|}{\textbf{Prediction:}} \\ 
        \hline
        C & $1.993 \pm 0.659$   & $0.125 \pm 0.150$   & $0.035 \pm 0.000$   \\ 
        \hline
        C & $-0.166 \pm 0.630$  & $2.271 \pm 0.620$   & $0.065 \pm 0.000$   \\ 
        \hline
        C & $0.533 \pm 0.561$   & $-1.742 \pm 0.843$  & $0.926 \pm 0.317$   \\ 
        \hline
        C & $-1.852 \pm 0.488$  & $0.700 \pm 0.359$   & $-0.212 \pm 0.250$  \\ 
        \hline
        H & $4.094 \pm 0.923$   & $-0.278 \pm 0.455$  & $-0.667 \pm 0.394$  \\ 
        \hline
        H & $0.494 \pm 1.370$   & $3.348 \pm 0.750$   & $-1.116 \pm 0.355$  \\ 
        \hline
        H & $-1.950 \pm 0.994$  & $-0.770 \pm 0.678$  & $1.688 \pm 0.498$   \\ 
        \hline
        H & $-1.950 \pm 0.703$  & $0.003 \pm 0.536$   & $-2.375 \pm 0.307$  \\ 
        \hline
        H & $1.994 \pm 1.201$   & $-3.170 \pm 0.735$  & $0.409 \pm 0.372$   \\ 
        \hline
        H & $-3.191 \pm 0.927$  & $-0.486 \pm 0.641$  & $1.248 \pm 0.488$   \\ 
        \hline
        \multicolumn{4}{|l|}{\textbf{Ground truth:}} \\ 
        \hline
        C & $1.652$      & $0.000$       & $0.000$       \\ 
        \hline
        C & $0.393$      & $2.362$       & $0.000$       \\ 
        \hline
        C & $0.687$      & $-2.209$      & $1.135$       \\ 
        \hline
        C & $-2.022$     & $1.147$       & $-0.529$      \\ 
        \hline
        H & $3.303$      & $-0.250$      & $-1.216$      \\ 
        \hline
        H & $1.093$      & $3.800$       & $-1.272$      \\ 
        \hline
        H & $-0.693$     & $-2.117$      & $2.636$       \\ 
        \hline
        H & $-2.596$     & $0.650$       & $-2.440$      \\ 
        \hline
        H & $1.542$      & $-4.022$      & $0.748$       \\ 
        \hline
        H & $-3.359$     & $0.639$       & $0.938$       \\ 
        \hline
    \end{tabular}
\end{supptable}

\begin{supptable}[h]
    \centering
    \caption{\textbf{Cyclobutane, S$_1$/S$_0$ Twisted Intersection}}
    \label{tab:c4h6_twisted_intersection}

    \begin{tabular}{|l|c|c|c|}
        \hline
        \textbf{Atom} & \textbf{X (a.u.)} & \textbf{Y (a.u.)} & \textbf{Z (a.u.)} \\ 
        \hline
        \multicolumn{4}{|l|}{\textbf{Prediction:}} \\ 
        \hline
        C & $2.129 \pm 0.453$  & $0.068 \pm 0.050$   & $-0.002 \pm 0.000$  \\ 
        \hline
        C & $2.070 \pm 1.002$  & $2.386 \pm 0.588$   & $0.114 \pm 0.100$   \\ 
        \hline
        C & $0.112 \pm 0.343$  & $-1.130 \pm 0.419$  & $0.646 \pm 0.226$   \\ 
        \hline
        C & $-1.892 \pm 0.562$ & $0.212 \pm 0.345$   & $-0.731 \pm 0.271$  \\ 
        \hline
        H & $2.552 \pm 0.759$  & $-0.732 \pm 0.390$  & $-1.845 \pm 0.373$  \\ 
        \hline
        H & $1.430 \pm 0.944$  & $2.121 \pm 0.798$   & $1.748 \pm 0.471$   \\ 
        \hline
        H & $-0.095 \pm 0.590$ & $-0.791 \pm 0.559$  & $2.242 \pm 0.356$   \\ 
        \hline
        H & $-2.069 \pm 0.783$ & $-0.281 \pm 0.567$  & $-2.559 \pm 0.458$  \\ 
        \hline
        H & $-0.405 \pm 0.561$ & $-3.044 \pm 0.307$  & $0.226 \pm 0.313$   \\ 
        \hline
        H & $-3.832 \pm 0.714$ & $1.193 \pm 0.389$   & $0.160 \pm 0.321$   \\ 
        \hline
        \multicolumn{4}{|l|}{\textbf{Ground truth:}} \\ 
        \hline
        C & $2.127$   & $-0.000$   & $0.000$    \\ 
        \hline
        C & $0.929$   & $2.462$    & $-0.000$   \\ 
        \hline
        C & $-0.112$  & $-1.601$   & $0.823$    \\ 
        \hline
        C & $-1.255$  & $0.437$    & $-0.898$   \\ 
        \hline
        H & $2.980$   & $-0.624$   & $-1.774$   \\ 
        \hline
        H & $0.396$   & $2.993$    & $1.921$    \\ 
        \hline
        H & $-0.679$  & $-1.327$   & $2.776$    \\ 
        \hline
        H & $-1.299$  & $-0.030$   & $-2.892$   \\ 
        \hline
        H & $-0.073$  & $-3.580$   & $0.269$    \\ 
        \hline
        H & $-3.016$  & $1.270$    & $-0.227$   \\ 
        \hline
    \end{tabular}
\end{supptable}

\begin{supptable}[h]
    \centering
    \caption{\textbf{Cyclobutane, S$_1$/S$_0$ Proton Migration}}
    \label{tab:c4h6_proton_migration}

    \begin{tabular}{|l|c|c|c|}
        \hline
        \textbf{Atom} & \textbf{X (a.u.)} & \textbf{Y (a.u.)} & \textbf{Z (a.u.)} \\ 
        \hline
        \multicolumn{4}{|l|}{\textbf{Prediction:}} \\ 
        \hline
        C & $3.277 \pm 0.869$  & $-0.016 \pm 0.050$ & $-0.007 \pm 0.000$  \\ 
        \hline
        H & $0.638 \pm 0.860$  & $3.211 \pm 0.364$  & $0.073 \pm 0.100$   \\ 
        \hline
        C & $0.972 \pm 0.642$  & $-1.496 \pm 0.882$ & $0.961 \pm 0.574$   \\ 
        \hline
        C & $0.653 \pm 0.456$  & $1.355 \pm 0.609$  & $-0.778 \pm 0.259$  \\ 
        \hline
        C & $-1.744 \pm 0.551$ & $0.030 \pm 0.243$  & $-0.004 \pm 0.278$  \\ 
        \hline
        H & $0.918 \pm 1.015$  & $-1.262 \pm 0.719$ & $2.969 \pm 0.358$   \\ 
        \hline
        H & $0.818 \pm 0.607$  & $-3.140 \pm 0.387$ & $-0.150 \pm 0.285$  \\ 
        \hline
        H & $0.623 \pm 0.801$  & $1.213 \pm 0.615$  & $-2.935 \pm 0.308$  \\ 
        \hline
        H & $-3.134 \pm 0.708$ & $1.163 \pm 0.507$  & $1.400 \pm 0.365$   \\ 
        \hline
        H & $-3.021 \pm 0.862$ & $-1.058 \pm 0.583$ & $-1.529 \pm 0.396$  \\ 
        \hline
        \multicolumn{4}{|l|}{\textbf{Ground truth:}} \\ 
        \hline
        C & $2.463$      & $-0.000$     & $-0.000$    \\ 
        \hline
        H & $0.637$      & $3.775$      & $-0.000$    \\ 
        \hline
        C & $0.657$      & $-1.945$     & $0.965$     \\ 
        \hline
        C & $0.657$      & $1.945$      & $-0.965$    \\ 
        \hline
        C & $-1.331$     & $0.000$      & $0.000$     \\ 
        \hline
        H & $0.637$      & $-2.284$     & $3.005$     \\ 
        \hline
        H & $0.637$      & $-3.775$     & $-0.000$    \\ 
        \hline
        H & $0.637$      & $2.284$      & $-3.005$    \\ 
        \hline
        H & $-2.498$     & $0.752$      & $1.517$     \\ 
        \hline
        H & $-2.498$     & $-0.752$     & $-1.517$    \\ 
        \hline
    \end{tabular}
\end{supptable}

\begin{supptable}[h]
    \centering
    \caption{\textbf{Cyclobutane, T = 605~fs}}
    \label{tab:c4h6_t_605}

    \begin{tabular}{|l|c|c|c|}
        \hline
        \textbf{Atom} & \textbf{X (a.u.)} & \textbf{Y (a.u.)} & \textbf{Z (a.u.)} \\ 
        \hline
        \multicolumn{4}{|l|}{\textbf{Prediction:}} \\ 
        \hline
        C & $2.213 \pm 0.592$  & $0.036 \pm 0.000$  & $-0.004 \pm 0.000$  \\ 
        \hline
        C & $-0.523 \pm 0.696$ & $1.986 \pm 0.550$  & $0.034 \pm 0.050$   \\ 
        \hline
        C & $0.639 \pm 0.343$  & $-1.833 \pm 0.563$ & $-0.047 \pm 0.153$  \\ 
        \hline
        C & $-1.543 \pm 0.675$ & $-0.657 \pm 0.268$ & $-0.547 \pm 0.196$  \\ 
        \hline
        H & $4.605 \pm 0.790$  & $0.183 \pm 0.320$  & $0.381 \pm 0.194$   \\ 
        \hline
        H & $1.161 \pm 1.136$  & $-3.825 \pm 0.536$ & $0.419 \pm 0.243$   \\ 
        \hline
        H & $0.948 \pm 1.323$  & $3.207 \pm 0.798$  & $-0.926 \pm 0.327$  \\ 
        \hline
        H & $-2.852 \pm 0.712$ & $0.777 \pm 0.573$  & $1.083 \pm 0.327$   \\ 
        \hline
        H & $-2.239 \pm 0.881$ & $1.773 \pm 0.752$  & $1.557 \pm 0.419$   \\ 
        \hline
        H & $-2.409 \pm 0.610$ & $-1.646 \pm 0.432$ & $-1.948 \pm 0.229$  \\ 
        \hline
        \multicolumn{4}{|l|}{\textbf{Ground truth:}} \\ 
        \hline
        C & $1.846$   & $0.000$    & $0.000$    \\ 
        \hline
        C & $0.399$   & $2.408$    & $0.000$    \\ 
        \hline
        C & $0.666$   & $-2.025$   & $0.132$    \\ 
        \hline
        C & $-2.036$  & $-1.048$   & $-0.382$   \\ 
        \hline
        H & $3.796$   & $0.175$    & $0.062$    \\ 
        \hline
        H & $1.230$   & $-3.902$   & $0.556$    \\ 
        \hline
        H & $1.509$   & $4.013$    & $-1.047$   \\ 
        \hline
        H & $-3.229$  & $-0.233$   & $1.076$    \\ 
        \hline
        H & $-0.893$  & $2.319$    & $1.343$    \\ 
        \hline
        H & $-3.288$  & $-1.707$   & $-1.740$   \\ 
        \hline
    \end{tabular}
\end{supptable}

\begin{supptable}[h]
    \centering
    \caption{\textbf{Cyclobutane, T = 615~fs}}
    \label{tab:c4h6_t_615}

    \begin{tabular}{|l|c|c|c|}
        \hline
        \textbf{Atom} & \textbf{X (a.u.)} & \textbf{Y (a.u.)} & \textbf{Z (a.u.)} \\ 
        \hline
        \multicolumn{4}{|l|}{\textbf{Prediction:}} \\ 
        \hline
        C & $2.495 \pm 0.538$  & $0.043 \pm 0.050$   & $-0.072 \pm 0.000$  \\ 
        \hline
        C & $-0.422 \pm 0.453$ & $1.810 \pm 0.586$   & $-0.056 \pm 0.000$  \\ 
        \hline
        C & $0.444 \pm 0.387$  & $-1.715 \pm 0.584$  & $0.099 \pm 0.204$   \\ 
        \hline
        C & $-2.333 \pm 0.499$ & $0.547 \pm 0.402$   & $0.366 \pm 0.205$   \\ 
        \hline
        H & $4.636 \pm 0.923$  & $0.461 \pm 0.367$   & $0.264 \pm 0.198$   \\ 
        \hline
        H & $0.478 \pm 1.365$  & $3.507 \pm 0.621$   & $-0.072 \pm 0.158$  \\ 
        \hline
        H & $2.428 \pm 1.100$  & $-2.611 \pm 0.608$  & $-0.368 \pm 0.235$  \\ 
        \hline
        H & $-2.255 \pm 0.869$ & $-1.450 \pm 0.601$  & $-0.985 \pm 0.290$  \\ 
        \hline
        H & $-1.866 \pm 0.904$ & $-2.249 \pm 0.570$  & $-1.031 \pm 0.276$  \\ 
        \hline
        H & $-3.605 \pm 0.874$ & $1.656 \pm 0.516$   & $1.855 \pm 0.319$   \\ 
        \hline
        \multicolumn{4}{|l|}{\textbf{Ground truth:}} \\ 
        \hline
        C & $1.851$      & $0.000$      & $0.000$      \\ 
        \hline
        C & $-0.062$     & $2.239$      & $0.000$      \\ 
        \hline
        C & $0.893$      & $-2.366$     & $0.007$      \\ 
        \hline
        C & $-2.376$     & $1.101$      & $0.447$      \\ 
        \hline
        H & $3.734$      & $0.437$      & $0.370$      \\ 
        \hline
        H & $0.810$      & $4.035$      & $0.036$      \\ 
        \hline
        H & $2.251$      & $-4.126$     & $-0.309$     \\ 
        \hline
        H & $-2.827$     & $-0.174$     & $-1.069$     \\ 
        \hline
        H & $-0.506$     & $-3.014$     & $-1.163$     \\ 
        \hline
        H & $-3.768$     & $1.867$      & $1.682$      \\ 
        \hline
    \end{tabular}
\end{supptable}

\begin{supptable}[h]
    \centering
    \caption{\textbf{Cyclobutane, T = 625~fs}}
    \label{tab:c4h6_t_625}

    \begin{tabular}{|l|c|c|c|}
        \hline
        \textbf{Atom} & \textbf{X (a.u.)} & \textbf{Y (a.u.)} & \textbf{Z (a.u.)} \\ 
        \hline
        \multicolumn{4}{|l|}{\textbf{Prediction:}} \\ 
        \hline
        C & $2.456 \pm 0.626$  & $0.139 \pm 0.100$  & $0.031 \pm 0.050$   \\ 
        \hline
        C & $1.593 \pm 1.018$  & $2.355 \pm 0.641$  & $-0.036 \pm 0.000$  \\ 
        \hline
        C & $0.114 \pm 0.629$  & $-1.843 \pm 0.979$ & $0.532 \pm 0.253$   \\ 
        \hline
        C & $-2.202 \pm 0.740$ & $-0.797 \pm 0.535$ & $0.450 \pm 0.212$   \\ 
        \hline
        H & $4.289 \pm 1.020$  & $-0.382 \pm 0.494$ & $-0.873 \pm 0.329$  \\ 
        \hline
        H & $-0.335 \pm 1.079$ & $-3.323 \pm 0.546$ & $-0.180 \pm 0.341$  \\ 
        \hline
        H & $2.927 \pm 1.466$  & $2.746 \pm 1.001$  & $-1.081 \pm 0.336$  \\ 
        \hline
        H & $-2.455 \pm 1.688$ & $2.205 \pm 0.866$  & $0.115 \pm 0.260$   \\ 
        \hline
        H & $-2.948 \pm 1.621$ & $1.561 \pm 0.725$  & $-0.112 \pm 0.290$  \\ 
        \hline
        H & $-3.441 \pm 1.684$ & $-2.661 \pm 0.592$ & $1.155 \pm 0.292$   \\ 
        \hline
        \multicolumn{4}{|l|}{\textbf{Ground truth:}} \\ 
        \hline
        C & $2.012$    & $0.000$     & $-0.000$    \\ 
        \hline
        C & $1.789$    & $2.712$     & $-0.000$    \\ 
        \hline
        C & $-0.199$   & $-2.002$    & $0.405$     \\ 
        \hline
        C & $-2.775$   & $-1.486$    & $0.610$     \\ 
        \hline
        H & $3.629$    & $-0.640$    & $-0.986$    \\ 
        \hline
        H & $-0.186$   & $-3.761$    & $-0.516$    \\ 
        \hline
        H & $3.345$    & $4.059$     & $-1.006$    \\ 
        \hline
        H & $-2.997$   & $0.621$     & $0.478$     \\ 
        \hline
        H & $-0.223$   & $3.316$     & $-0.119$    \\ 
        \hline
        H & $-4.396$   & $-2.819$    & $1.133$     \\ 
        \hline
    \end{tabular}
\end{supptable}

\clearpage
\subsection*{Supplementary Note 1 - Model Details}
\subsubsection*{Embedding Module}
The raw input to MOLEXA consists of the atomic number \( z_i \), charge state \( q_i \), and momentum \( \mathbf{p}_i \) of all ion fragments produced by the Coulomb explosion of a molecule. The Embedding Module (Algorithm~\ref{alg:EmbeddingModule}) converts the atomic number and charge state to embeddings and concatenates them with the linearly transformed momentum features by using the Input Embedder (Algorithm~\ref{alg:InputEmbedder}). The resulting features of each atomic pair in the molecule are further concatenated to form the pairwise features, which are processed by the Pair Residual Block (Algorithm~\ref{alg:PairResidualBlock}) before being sent to the Dynamics Extraction Module (Algorithm~\ref{alg:DynamicsExtractionModule}). The Pair Residual Block is a two-layer perceptron with a residual connection.

\begin{algorithm}[H]
\caption{Embedding Module}\label{alg:EmbeddingModule}
\begin{algorithmic}[1]
\STATE \textbf{Function} \textsc{EmbeddingModule}$\big(\{z_i\},\{q_i\}, \{\mathbf{p}_i\}\big):$
\STATE \quad $\{\mathbf{a}_i\} = \hyperref[alg:InputEmbedder]{\text{InputEmbedder}}(\{z_i\}, \{q_i\}, \{\mathbf{p}_i\})$ \hfill \textit{\small $\mathbf{a}_i \in \mathbb{R}^{c_a}, \; c_a = 192$}
\STATE \quad $\mathbf{e}_{ij} = \text{Concat}([\mathbf{a}_i, \mathbf{a}_j])$
\STATE \quad $\{\mathbf{b}_{ij}\} \gets \hyperref[alg:PairResidualBlock]{\text{PairResidualBlock}}(\{\mathbf{e}_{ij}\}, c_{br} = 384)$  \hfill \textit{\small $\mathbf{b}_{ij} \in \mathbb{R}^{c_b}, \; c_b = 384$}
\STATE \quad \textbf{return} $\{\mathbf{b}_{ij}\}$
\end{algorithmic}
\end{algorithm}

\begin{algorithm}[H]
\caption{Input Embedder}\label{alg:InputEmbedder}
\begin{algorithmic}[1]
\STATE \textbf{Function} \textsc{InputEmbedder}$\big(\{z_i\},\{q_i\}, \{\mathbf{p}_i\}\big):$
\STATE \quad $\{\mathbf{r}_i\} = \text{Embedding}(\{z_i\})$ \hfill \textit{\small $\mathbf{r}_i \in \mathbb{R}^{c}, \; c = 64$}
\STATE \quad $\{\mathbf{s}_i\} = \text{Embedding}(\{q_i\})$ \hfill \textit{\small $\mathbf{s}_i \in \mathbb{R}^{c}, \; c = 64$}
\STATE \quad $\mathbf{t}_i = \text{Linear}(\mathbf{p}_i)$ \hfill \textit{\small $\mathbf{t}_i \in \mathbb{R}^{c}, \; c = 64$}
\STATE \quad $\mathbf{a}_{i} = \text{Concat}([\mathbf{r}_i, \mathbf{s}_i, \mathbf{t}_i])$
\STATE \quad \textbf{return} $\{\mathbf{a}_{i}\}$
\end{algorithmic}
\end{algorithm}

\begin{algorithm}[H]
\caption{Pair Residual Block}\label{alg:PairResidualBlock}
\begin{algorithmic}[1]
\STATE \textbf{Function} \textsc{PairResidualBlock}$\big(\{\mathbf{b}_{ij}\}, c_{br}, Activation1 = ReLU, Activation2 = ReLU\big):$
\STATE \quad $\mathbf{a}_{ij} = \text{Activation1}(\text{Linear}(\mathbf{b}_{ij}))$ \hfill \textit{\small $\mathbf{a}_{ij} \in \mathbb{R}^{c_{br}}$}
\STATE \quad $\mathbf{d}_{ij} = \text{Activation2}(\text{Linear}(\mathbf{a}_{ij}))$ \hfill \textit{\small $\mathbf{d}_{ij} \in \mathbb{R}^{c_{br}}$}
\STATE \quad $\mathbf{b}_{ij} \gets \text{LayerNorm}(\mathbf{b}_{ij} + \mathbf{d}_{ij})$
\STATE \quad \textbf{return} $\{\mathbf{b}_{ij}\}$
\end{algorithmic}
\end{algorithm}

\subsubsection*{Dynamics Extraction Module}
The Dynamics Extraction Module (Algorithm~\ref{alg:DynamicsExtractionModule}) is illustrated in Fig.~\ref{fig:illustration_and_model}. Its task is to generate the dynamics-specific conditions to be used in the Structure Denoising Module (Algorithm~\ref{alg:StructureDenoisingModule}). It does this by processing the pairwise features with six consecutive TM blocks (Algorithm~\ref{alg:TransformerWithMemory}). In each of the six blocks, the features are first transformed based on the inter-pair correlations with the pairwise Multi-head Self-attention block (Algorithm~\ref{alg:PairAttention}). They are then passed to a Pair Residual Block (Algorithm~\ref{alg:PairResidualBlock}). The subsequent memory operations regulate the information to be passed to the next TM block.

\begin{algorithm}
\caption{Dynamics Extraction Module}\label{alg:DynamicsExtractionModule}
\begin{algorithmic}[1]
\STATE \textbf{Function} \textsc{DynamicsExtractionModule}$\big(\{\mathbf{b}_{ij}\}, N_{block} = 6\big):$
\STATE \quad $\{\mathbf{m}_{ij}\} \gets \{\mathbf{b}_{ij}\}$
\STATE \quad \textbf{for} $n \in [1, \dots, N_\text{block}]$ \textbf{do}
\STATE \quad \quad $\{\mathbf{b}_{ij}\}, \{\mathbf{m}_{ij}\} \gets \hyperref[alg:TransformerWithMemory]{\text{TransformerWithMemory}}(\{\mathbf{b}_{ij}\}, \{\mathbf{m}_{ij}\})$
\STATE \quad \textbf{end for}
\STATE \quad \textbf{return} $\{\mathbf{b}_{ij}\}$
\end{algorithmic}
\end{algorithm}

\begin{algorithm}
\caption{Transformer with Memory}\label{alg:TransformerWithMemory}
\begin{algorithmic}[1]
\STATE \textbf{Function} \textsc{TransformerWithMemory}$\big(\{\mathbf{b}_{ij}\}, \{\mathbf{m}_{ij}\}\big):$
\STATE \quad \textit{\textcolor{purple}{\# Attention block}}
\STATE \quad $\{\mathbf{e}_{ij}\} = \hyperref[alg:PairAttention]{\text{PairAttention}}(\{\mathbf{b}_{ij}\})$
\STATE \quad $\mathbf{e}_{ij} \gets \text{LayerNorm}(\mathbf{e}_{ij} + \mathbf{b}_{ij})$
\STATE \quad $\{\mathbf{b}_{ij}\} \gets \hyperref[alg:PairResidualBlock]{\text{PairResidualBlock}}(\{\mathbf{e}_{ij}\}, c_{br} = 384)$  \hfill \textit{\small $\mathbf{b}_{ij} \in \mathbb{R}^{c_b}, \; c_b = 384$}
\STATE \quad \textit{\textcolor{purple}{\# Memory block}}
\STATE \quad $\mathbf{\tilde{m}}_{ij} = \text{tanh}(\text{Linear}(\mathbf{b}_{ij}))$ \hfill \textit{\small $\mathbf{\tilde{m}}_{ij} \in \mathbb{R}^{c_b}, \; c_b = 384$}
\STATE \quad $\mathbf{u}_{ij}, \mathbf{f}_{ij}, \mathbf{o}_{ij} = \text{sigmoid}(\text{Linear}(\mathbf{b}_{ij}))$ \hfill \textit{\small $\mathbf{u}_{ij}, \mathbf{f}_{ij}, \mathbf{o}_{ij} \in \mathbb{R}^{c_b}, \; c_b = 384$}
\STATE \quad $\mathbf{m}_{ij} = \mathbf{u}_{ij} \odot \mathbf{\tilde{m}}_{ij} + \mathbf{f}_{ij} \odot \mathbf{m}_{ij}$
\STATE \quad $\mathbf{b}_{ij} = \mathbf{o}_{ij} \odot \text{tanh}(\mathbf{m}_{ij})$
\STATE \quad \textbf{return} $\{\mathbf{b}_{ij}\}$, $\{\mathbf{m}_{ij}\}$
\end{algorithmic}
\end{algorithm}


\begin{algorithm}[H]
\caption{Pairwise Multi-head Self-attention}\label{alg:PairAttention}
\begin{algorithmic}[1]
\STATE \textbf{Function} \textsc{PairAttention}$\big(\{\mathbf{b}_{ij}\}, N_{head} = 32\big):$
\STATE \quad $\mathbf{q}^h_g, \mathbf{k}^h_g, \mathbf{v}^h_g = \text{LinearNoBias}(\mathbf{b}_{ij})$ \hfill \textit{\small $ h \in \{1, \dots, N_{head}\}, \mathbf{q}^h_g, \mathbf{k}^h_g, \mathbf{v}^h_g \in \mathbb{R}^{c_h}, \; c_h = 12$}
\STATE \quad $\mathbf{w}^h_{gl} = \text{softmax}\Big(\frac{\mathbf{q}^{h}_{g}\mathbf{k}^{h^\top}_{l}}{\sqrt{c_h}}\Big)$ 
\STATE \quad $\mathbf{y}^h_g = \sum_l \mathbf{w}^h_{gl}\mathbf{v}^h_l$ 
\STATE \quad $\mathbf{b}_{ij} \gets \text{LinearNoBias}(\mathbf{y}^h_g)$ \hfill \textit{\small $\mathbf{b}_{ij} \in \mathbb{R}^{c_b}, \; c_b = 384$}
\STATE \quad \textbf{return} $\{\mathbf{b}_{ij}\}$
\end{algorithmic}
\end{algorithm}

\subsubsection*{Structure Denoising Module}
In the Structure Denoising Module (Algorithm~\ref{alg:StructureDenoisingModule}), a noisified (training) or random (inference) molecular structure is conditionally encoded by the Conditional Position Encoder (Algorithm~\ref{alg:ConditionalEncoder}). The conditions are the noise level \( \sigma \) and the pairwise features produced by the Dynamics Extraction Module (Algorithm~\ref{alg:DynamicsExtractionModule}). The former is Fourier encoded, linearly transformed, and then concatenated with the linearly transformed positions. The result is used to create the pairwise features which are passed together with the dynamics-specific conditions to a Pair Residual Block (Algorithm~\ref{alg:PairResidualBlock}). The produced encoding is processed by two TM blocks (Algorithm~\ref{alg:TransformerWithMemory}). Atom-wise features are then created through projection of the calculated pairwise features. They are processed by an Atom-wise Multi-head Self-attention block (Algorithm~\ref{alg:AtomAttention}), with the result serving as input to the Position Decoder (Algorithm~\ref{alg:PositionDecoder}). The decoding is performed with an Atom Residual block (Algorithm~\ref{alg:AtomResidualBlock}), which is a two-layer perceptron without nonlinear activation at the end. The final output is a weighted sum of the decoded structure and the input molecular structure. To predict accurate molecular structures, the Structure Denoising Module is run by a diffusion sampler (Algorithm~\ref{alg:Sampler}) through five iterations. The sampler is based on the approach proposed in Ref.~\cite{karras_elucidating_2022}. Its parameter settings were optimized with the validation dataset. The prediction errors evaluated as a function of the number of sampling steps (NSS) are displayed in Supplementary Fig.~\ref{fig:S_err_nfe}. The overall minimum is reached at the NSS of 33. Since five sampling steps can already get to an error within 1\% of the minimum value (achievable with less than 100 sampling steps), it was chosen as a good compromise between prediction accuracy and inference time.

\begin{algorithm}
\caption{Structure Denoising Module}\label{alg:StructureDenoisingModule}
\begin{algorithmic}[1]
\STATE \textbf{Function} \textsc{StructureDenoisingModule}$\big(\mathbf{\sigma}, \{\mathbf{x}_i\}, \{\mathbf{b}_{ij}\}, N_{block} = 2\big):$
\STATE \quad $\{\mathbf{b}_{ij}\} \gets \hyperref[alg:ConditionalEncoder]{\text{ConditionalEncoder}}(\mathbf{\sigma}, \{\mathbf{x}_i\}, \{\mathbf{b}_{ij}\})$ \hfill \textit{\small $\mathbf{b}_{ij} \in \mathbb{R}^{c_b}, \; c_b = 384$}
\STATE \quad $\{\mathbf{m}_{ij}\} \gets \{\mathbf{b}_{ij}\}$
\STATE \quad \textbf{for} $n \in [1, \dots, N_\text{block}]$ \textbf{do}
\STATE \quad \quad $\{\mathbf{b}_{ij}\}, \{\mathbf{m}_{ij}\} \gets \hyperref[alg:TransformerWithMemory]{\text{TransformerWithMemory}}(\{\mathbf{b}_{ij}\}, \{\mathbf{m}_{ij}\})$
\STATE \quad \textbf{end for}
\STATE \quad $\mathbf{a}_{i} = \sum_j\mathbf{b}_{ij}$
\STATE \quad $\mathbf{a}_{i} \gets \text{LayerNorm}(\mathbf{a}_{i})$
\STATE \quad $\{\mathbf{d}_{i}\} = \hyperref[alg:AtomAttention]{\text{AtomAttention}}(\{\mathbf{a}_{i}\})$
\STATE \quad $\mathbf{a}_{i} \gets \text{LayerNorm}(\mathbf{a}_{i} + \mathbf{d}_{i})$
\STATE \quad $\{\mathbf{x}_{i}\} \gets \hyperref[alg:PositionDecoder]{\text{PositionDecoder}}(\mathbf{\sigma}, \{\mathbf{x}_i\}, \{\mathbf{a}_i\})$  \hfill \textit{\small $\mathbf{x}_{i} \in \mathbb{R}^{c_o}, \; c_o = 3$}
\STATE \quad \textbf{return} $\{\mathbf{x}_{i}\}$
\end{algorithmic}
\end{algorithm}

\begin{algorithm}
\caption{Conditional Position Encoder}\label{alg:ConditionalEncoder}
\begin{algorithmic}[1]
\STATE \textbf{Function} \textsc{ConditionalEncoder}$\big(\mathbf{\sigma}, \{\mathbf{x}_i\},\{\mathbf{b}_{ij}\}, \sigma_{d} = 0.25\}\big):$
\STATE \quad \textit{\textcolor{purple}{\# Embedding noise levels}}
\STATE \quad $\mathbf{f} \sim \mathcal{N}(\mathbf{0}, \mathbf{I})$ \hfill \textit{\small $\mathbf{f} \in \mathbb{R}^{c_f}, \; c_f = 128$}
\STATE \quad $\mathbf{\sigma} \gets 8\pi \times \text{log}(\mathbf{\sigma}) \otimes \mathbf{f}$
\STATE \quad $\mathbf{\sigma} \gets \text{Concat}([\text{cos}(\mathbf{\sigma}), \text{sin}(\mathbf{\sigma})])$
\STATE \quad $\mathbf{s} = \text{Linear}(\mathbf{\sigma})$ \hfill \textit{\small $\mathbf{s} \in \mathbb{R}^{c_s}, \; c_s = 256$}
\STATE \quad \textit{\textcolor{purple}{\# Combine noise and structure features}}
\STATE \quad $c_{in} = \sqrt{\frac{1}{\sigma_{d}^{\ 2}\ +\ \sigma^2}}$ 
\STATE \quad $\mathbf{x}_i \gets c_{in}\mathbf{x}_i$ 
\STATE \quad $\mathbf{y}_i = \text{Linear}(\mathbf{x}_i)$ \hfill \textit{\small $\mathbf{y}_i \in \mathbb{R}^{c_s}, \; c_s = 256$}
\STATE \quad $\mathbf{x}_i \gets \text{Concat}([\mathbf{s}, \mathbf{y}_i])$
\STATE \quad $\mathbf{x}_{ij} \gets \text{Concat}([\mathbf{x}_i, \mathbf{x}_j])$
\STATE \quad \textit{\textcolor{purple}{\# Condition the combined features on the extracted dynamics features}}
\STATE \quad $\mathbf{x}_{ij} \gets \text{Concat}([\mathbf{b}_{ij}, \mathbf{x}_{ij}])$
\STATE \quad $\{\mathbf{b}_{ij}\} \gets \hyperref[alg:PairResidualBlock]{\text{PairResidualBlock}}(\{\mathbf{x}_{ij}\}, c_{br} = 384)$  \hfill \textit{\small $\mathbf{b}_{ij} \in \mathbb{R}^{c_b}, \; c_b = 384$}
\STATE \quad \textbf{return} $\{\mathbf{b}_{ij}\}$
\end{algorithmic}
\end{algorithm}

\clearpage

\begin{algorithm}
\caption{Atom-wise Multi-head Self-attention}\label{alg:AtomAttention}
\begin{algorithmic}[1]
\STATE \textbf{Function} \textsc{AtomAttention}$\big(\{\mathbf{a}_{i}\}, N_{head} = 32\big):$
\STATE \quad $\mathbf{q}^h_i, \mathbf{k}^h_i, \mathbf{v}^h_i = \text{LinearNoBias}(\mathbf{a}_{i})$ \hfill \textit{\small $ h \in \{1, \dots, N_{head}\}, \mathbf{q}^h_i, \mathbf{k}^h_i, \mathbf{v}^h_i \in \mathbb{R}^{c_h}, \; c_h = 12$}
\STATE \quad $\mathbf{w}^h_{ij} = \text{softmax}\Big(\frac{\mathbf{q}^{h}_{i}\mathbf{k}^{h^\top}_{j}}{\sqrt{c_h}}\Big)$ 
\STATE \quad $\mathbf{y}^h_i = \sum_j \mathbf{w}^h_{ij}\mathbf{v}^h_j$ 
\STATE \quad $\mathbf{a}_{i} \gets \text{LinearNoBias}(\mathbf{y}^h_i)$ \hfill \textit{\small $\mathbf{a}_{i} \in \mathbb{R}^{c_b}, \; c_b = 384$}
\STATE \quad \textbf{return} $\{\mathbf{a}_{i}\}$
\end{algorithmic}
\end{algorithm}

\begin{algorithm}
\caption{Position Decoder}\label{alg:PositionDecoder}
\begin{algorithmic}[1]
\STATE \textbf{Function} \textsc{PositionDecoder}$\big(\mathbf{\sigma}, \{\mathbf{x}_i\}, \{\mathbf{a}_i\}, \sigma_{d} = 0.25\}\big):$
\STATE \quad $c_{skip} = \frac{\sigma_{d}^{\ 2}}{\sigma_{d}^{\ 2}\ +\ \sigma^2}$ 
\STATE \quad $c_{out} = \sqrt{\frac{\sigma_{d}\sigma}{\sigma_{d}^{\ 2}\ +\ \sigma^2}}$ 
\STATE \quad $\{\mathbf{y}_{i}\} \gets \hyperref[alg:AtomResidualBlock]{\text{AtomResidualBlock}}(\{\mathbf{a}_{i}\}, c_{ar} = 3)$  \hfill \textit{\small $\mathbf{y}_{i} \in \mathbb{R}^{c_o}, \; c_o = 3$}
\STATE \quad $\mathbf{x}_{i} \gets c_{skip}\mathbf{x}_{i} + c_{out}\mathbf{y}_{i}$
\STATE \quad \textbf{return} $\{\mathbf{x}_{i}\}$
\end{algorithmic}
\end{algorithm}

\begin{algorithm}
\caption{Atom Residual Block}\label{alg:AtomResidualBlock}
\begin{algorithmic}[1]
\STATE \textbf{Function} \textsc{AtomResidualBlock}$\big(\{\mathbf{a}_{i}\}, c_{ar}, Activation1 = ReLU, Activation2 = ReLU\big):$
\STATE \quad $\mathbf{x}_{i} = \text{Activation1}(\text{Linear}(\mathbf{a}_{i}))$ \hfill \textit{\small $\mathbf{x}_{i} \in \mathbb{R}^{c_{ar}}$}
\STATE \quad $\mathbf{y}_{i} = \text{Linear}(\mathbf{x}_{i})$ \hfill \textit{\small $\mathbf{y}_{i} \in \mathbb{R}^{c_{ar}}$}
\STATE \quad $\mathbf{y}_{i} \gets \mathbf{a}_{i} + \mathbf{y}_{i}$
\STATE \quad \textbf{return} $\{\mathbf{y}_{i}\}$
\end{algorithmic}
\end{algorithm}

\clearpage
\begin{algorithm}[H]
\caption{Diffusion Sampler}\label{alg:Sampler}
\begin{algorithmic}[1]
\STATE \textbf{Function} \textsc{Sampler}$\big(\{\mathbf{b}_{ij}\},\sigma_{min} = 0.002, \sigma_{max} = 80, \rho = 1.5, S_{churn} = 30, S_{min} = 0.01, S_{max} = 1, S_{noise} = 1.1, N_{step} = 5\big):$
\STATE \quad $\mathbf{t} = [t_0, \dots, t_i,\dots,t_{N_{step}-1},0]$ \hfill \textit{\small $t_i = (\sigma_{max}^{\frac{1}{\rho}} + \frac{i}{(N_{step} - 1)(\sigma_{min}^{\frac{1}{\rho}} - \sigma_{max}^{\frac{1}{\rho}})})^{\rho}$}
\STATE \quad $\mathbf{x}_{n+1} \sim \mathcal{N}(\mathbf{0}, \mathbf{I})$ \hfill \textit{\small $\mathbf{x}_{n+1} \in \mathbb{R}^{c_o}, \; c_o = 3$}
\STATE \quad $\mathbf{x}_{n+1} \gets \mathbf{x}_{n+1} \times t_0$
\STATE \quad \textbf{for} $n \in [0, \dots, N_\text{step}-1]$ \textbf{do}
\STATE \quad \quad $\mathbf{x}_{n} \gets \mathbf{x}_{n+1}$
\STATE \quad \quad \textbf{if} $S_{min} <= t_{n-1} <= S_{max}$ 
\STATE \quad \quad \quad$\gamma = S_{churn}$
\STATE \quad \quad \textbf{else}
\STATE \quad \quad \quad$\gamma = 0$
\STATE \quad \quad \textbf{end if}
\STATE \quad \quad $\hat{t} = t_{n}(1 + \gamma)$
\STATE \quad \quad $\mathbf{d} \sim \mathcal{N}(\mathbf{0}, \mathbf{I})$ \hfill \textit{\small $\mathbf{d} \in \mathbb{R}^{c_o}, \; c_o = 3$}
\STATE \quad \quad $\hat{\mathbf{x}} = \mathbf{x}_{n} + \mathbf{d} \times S_{noise}\sqrt{\hat{t}^2 - t_{n}^2}$
\STATE \quad \quad $\{\mathbf{y}\} = \hyperref[alg:StructureDenoisingModule]{\text{StructureDenoisingModule}}(\hat{t}, \{\hat{\mathbf{x}}\}, \{\mathbf{b}_{ij}\})$
\STATE \quad \quad $\mathbf{\delta} = \frac{\hat{\mathbf{x}} - \mathbf{y}}{\hat{t}}$
\STATE \quad \quad $\mathbf{x}_{n+1} \gets \hat{\mathbf{x}} + (t_{n+1} - \hat{t})\mathbf{\delta}$
\STATE \quad \textbf{return} $\{\mathbf{x}_{n+1}\}$
\end{algorithmic}
\end{algorithm}

\clearpage

\subsubsection*{Uncertainty Estimation Module}
The Uncertainty Estimation Module (Algorithm~\ref{alg:UncertaintyModule}) pre-defines the uncertainty bins \( \mathbf{r} = [0, 0.05, 0.1, \dots, 9.95] \) and estimates the probability that the prediction error belonging to each of these bins. The input to the Uncertainty Estimation Module is the reconstructed molecular structure from the final sampling step and the pairwise features produced by the Dynamics Extraction Module (Algorithm~\ref{alg:DynamicsExtractionModule}). Similarly to the Structure Denoising Module, the reconstructed molecular structure is conditionally encoded with the pairwise features through a Pair Residual block (Algorithm~\ref{alg:PairResidualBlock}). The encoding is sent to two TM blocks (Algorithm~\ref{alg:TransformerWithMemory}) and then projected as atom-wise features. The probabilities for the 200 uncertainty bins are then predicted with an Atom Residual block (Algorithm~\ref{alg:AtomResidualBlock}) followed by a softmax layer. For each predicted coordinate, the uncertainty is then calculated as the probability-weighted sum of the bin values. Such values are directly used in Supplementary Tables~\ref{tab:h2o}-\ref{tab:c2h6o}. The uncertainty used in Fig.~\ref{fig:performance} and Supplementary Figs.~\ref{fig:S_accuracy_mae_uncertainty}-\ref{fig:S_max_DE_AE} was calculated by taking an average of the uncertainty values predicted for all atomic coordinates in a molecule.

\begin{algorithm}
\caption{Uncertainty Estimation Module}\label{alg:UncertaintyModule}
\begin{algorithmic}[1]
\STATE \textbf{Function} \textsc{UncertaintyModule}$\big(\{\mathbf{x}_i\},\{\mathbf{b}_{ij}\}, \mathbf{r} = [0, 0.05, 1, \dots, 9.95], N_{block} = 2\big):$
\STATE \quad $\mathbf{y}_i = \text{Linear}(\mathbf{x}_i)$ \hfill \textit{\small $\mathbf{y}_i \in \mathbb{R}^{c_s}, \; c_s = 256$}
\STATE \quad $\mathbf{y}_{ij} \gets \text{Concat}([\mathbf{y}_i, \mathbf{y}_j])$
\STATE \quad $\mathbf{x}_{ij} \gets \text{Concat}([\mathbf{y}_{ij}, \mathbf{b}_{ij}])$
\STATE \quad $\{\mathbf{b}_{ij}\} \gets \hyperref[alg:PairResidualBlock]{\text{PairResidualBlock}}(\{\mathbf{x}_{ij}\}, c_{br} = 384)$  \hfill \textit{\small $\mathbf{b}_{ij} \in \mathbb{R}^{c_b}, \; c_b = 384$}
\STATE \quad $\{\mathbf{m}_{ij}\} \gets \{\mathbf{b}_{ij}\}$
\STATE \quad \textbf{for} $n \in [1, \dots, N_\text{block}]$ \textbf{do}
\STATE \quad \quad $\{\mathbf{b}_{ij}\}, \{\mathbf{m}_{ij}\} \gets \hyperref[alg:TransformerWithMemory]{\text{TransformerWithMemory}}(\{\mathbf{b}_{ij}\}, \{\mathbf{m}_{ij}\})$
\STATE \quad \textbf{end for}
\STATE \quad $\mathbf{a}_{i} = \sum_j\mathbf{b}_{ij}$
\STATE \quad $\mathbf{a}_{i} \gets \text{LayerNorm}(\mathbf{a}_{i})$
\STATE \quad $\{\mathbf{s}_{i}\} \gets \hyperref[alg:AtomResidualBlock]{\text{AtomResidualBlock}}(\{\mathbf{a}_{i}\}, c_{ar} = 200)$  \hfill \textit{\small $\mathbf{s}_{i} \in \mathbb{R}^{c_{s}}, \; c_{s} = 20$}
\STATE \quad $\mathbf{s}_{i} \gets \text{softmax}(\mathbf{s}_{i})$
\STATE \quad $\mathbf{t} = \sum_i \mathbf{r}_{i}\mathbf{s}_i$ 
\STATE \quad \textbf{return} $\{\mathbf{t}\}$
\end{algorithmic}
\end{algorithm}

\end{document}